\documentclass{iopart}

\usepackage[dvips]{graphicx}

\begin{document}

\title[Fermi-normal, optical, and wave-synchronous
  coordinates]{Fermi-normal, optical, and wave-synchronous coordinates
  for spacetime with a plane gravitational wave}

\author{Malik~Rakhmanov}

\address{Center for Gravitational Wave Astronomy, Department of
  Physics, University of Texas at Brownsville, 1 W University Blvd.,
  Brownsville, TX 78520, USA}

\ead{malik.rakhmanov@utb.edu}

\begin{abstract}
Fermi normal coordinates provide a standardized way to describe 
the effects of gravitation from the point of view of an inertial
observer. These coordinates have always been introduced via
perturbation expansions and were usually limited to distances much
less than the characteristic length scale set by the curvature of
spacetime. For a plane gravitational wave this scale is given by its
wavelength which defines the domain of validity for these coordinates
known as the long-wavelength regime. The symmetry of this spacetime,
however, allows us to extend Fermi normal coordinates far beyond the
long-wavelength regime. Here we present an explicit construction for
this long-range Fermi normal coordinate system based on the unique
solution of the boundary-value problem for spacelike geodesics. The
resulting formulae amount to summation of the infinite series for
Fermi normal coordinates previously obtained with perturbation
expansions. We also consider two closely related normal coordinate
systems: optical coordinates which are built from null geodesics and
wave-synchronous coordinates which are built from spacelike geodesics
locked in phase with the propagating gravitational wave. The
wave-synchronous coordinates yield the exact solution of Peres and
Ehlers-Kundt which is globally defined. In this case, the limitation 
of the long-wavelength regime is completely overcome, and the system
of wave-synchronous coordinates becomes valid for arbitrarily large 
distances. Comparison of the different coordinate systems is done by
considering the motion of an inertial test mass in the field of a 
plane gravitational wave. 
\end{abstract}


\section{Introduction}
\label{sect:Intro}

In general relativity the choice of coordinates is rather arbitrary 
and no preference is given to any particular coordinate system ahead
of time. However, when the observer wants to describe the effects of
gravitation in his vicinity, he may find it convenient to use a
quasi-Cartesian coordinate system also known as the local-Lorentz
frame. This coordinate system is associated with a reference point in 
spacetime usually chosen at its origin. A continuous set of 
quasi-Cartesian (QC) coordinates associated with a reference curve 
gives rise to what is known as Fermi coordinates 
\cite{Fermi:1922,O'Raifeartaigh:1958}, if a certain condition 
is satisfied. Namely, to guarantee uniqueness, one must preclude 
arbitrary rotations of the QC frame, allowing only the rotation  
which is caused by the bending of the reference curve. The basis 
vectors (tetrad) which define the orientation of these coordinates are 
carried along the curve by means of Fermi-Walker transport. From a 
mathematical point of view, the Fermi coordinates represent a unique, 
continuous set of non-rotating QC coordinates, in which the metric is 
flat and all of its first derivatives vanish on the reference curve,
except perhaps the derivative along the curve 
\cite{Fermi:1922,O'Raifeartaigh:1958}. From a physical point of view, 
these are the coordinates that the observer would naturally use to
measure distances and times in his vicinity, whereas the reference
curve is his worldline in the four-dimensional spacetime
\cite{Synge:1960}. If the observer is freely falling, the reference
curve becomes geodesic and the Fermi-Walker transport of the basis
vectors becomes parallel transport. In this case, all the first
derivatives of the metric vanish on the reference curve and the
resulting coordinates are called {\emph{Fermi normal}} (FN)
\cite{Manasse:1963}. For an accelerating observer the definition of
Fermi coordinates is somewhat more complicated but is introduced along
the same lines (e.g. \cite{Ni:1978}). Note that although an observer 
on Earth is not inertial, one often ignores this fact to simplify
calculations. In particular, the effect of gravitational waves on a
measuring device (detector) in a laboratory environment on Earth is
often described from the point of view of such an inertial observer,
neglecting the gravitational field of Earth.

One of the earliest descriptions of a gravitational wave interacting  
with a detector as viewed by an inertial observer was introduced by
Weber \cite{Weber:1961}. This picture was widely used at the time when 
resonant bar detectors were operating around the Earth in search of
cosmic gravitational waves. It is a curious fact that this approach 
is largely forgotten today. In modern times of laser-interferometric
gravitational wave detectors, the coordinates of choice are those in
which a gravitational wave is described by the transverse and
traceless tensor and which are often referred to as 
{\emph{the TT coordinates}} or the TT gauge \cite{Misner:1973}. In these 
coordinates, one can calculate the response of laser interferometers 
to gravitational waves with relative ease and with no limit on the
distances spanned by these coordinates (e.g. \cite{Rakhmanov:2008}).  
Similar calculations in FN coordinates are usually more complicated 
and thus far have always been restricted to distances much less than 
the wavelength of the gravitational wave, the condition commonly known 
as {\emph{the long-wavelength regime.}} This is mainly why FN
coordinates have been gradually displaced by TT coordinates over time.

Historically, the local coordinates associated with an observer have
been introduced within different mathematical frameworks and appeared 
under different names before a standardized approach emerged. In the
early days, the coordinate construction would simply be an adaptation
of the geodesic deviation equation (e.g. \cite{Weber:1961,Misner:1973})
and, as a result, it was naturally limited to the long-wavelength
regime. Pertaining to the center of mass of a resonant bar, such a
coordinate system was frequently referred to as the rest frame of the 
detector. In laser interferometers, a similar coordinate system would 
be associated with the interferometer beam splitter and therefore
would be called the rest frame of the beam splitter. Early analysis 
of gravitational waves in these coordinates can be found in papers by
Grishchuk \cite{Grishchuk:1977}, Grishchuk and Polnaver
\cite{Grishchuk:1980}, and Pegoraro \etal \cite{Pegoraro:1978}. 
In all these cases, the calculations were carried out under the
assumption that the distances spanned by these coordinates are much
less than the wavelength of the gravitational wave. A more direct
approach to the detector coordinates was taken by Fortini and Gualdi 
\cite{Fortini:1982} who chose the Fermi normal construction as their 
main tool, cutting short the equation for geodesic deviation. 
This approach was later adopted by others and gradually became the
method of choice when the analysis of gravitational waves had to be
carried out from the point of view of an inertial observer 
\cite{Flores:1986,Callegari:1987,Fortini:1990,Fortini:1991}. From 
then on, there was no need to consider closely located geodesics, 
and yet the distance limitation inherited from the equation for
geodesic deviation continued to appear in all subsequent calculations 
with these coordinates.

More recent interest in the coordinate system of an inertial observer 
was motivated by the desire to understand better the response of 
interferometric gravitational-wave detectors. Two different efforts 
were made to extend the detector response beyond the long-wavelength 
regime in the local-Lorentz frame \cite{Baskaran:2004, Rakhmanov:2005}. 
The first targeted linear-frequency corrections to the 
long-wavelength approximation, whereas the second attempted to obtain 
formulae to all orders of the perturbation expansion. It was soon 
realized that the coordinate systems used in these calculations were 
different even though both approaches were based on the local-Lorentz 
frame of an inertial observer. The difference in the coordinate
systems was the initial motivation for this paper. Another motivation 
came from the fact that the coordinates utilized in \cite{Rakhmanov:2005} 
did not have the usual higher-order corrections. By chance, the metric
in these coordinates turned out to be equivalent to the exact solution 
of Peres \cite{Peres:1959} and Ehlers-Kundt \cite{Ehlers:1962}, 
which is globally defined. Different questions naturally appear 
at this point. What is the relationship between these new coordinates 
and the Fermi normal frame? And, why is there even an
ambiguity in the definition of a normal-coordinate system?

To answer these questions we have to consider the Fermi normal
construction in its full generality, i.e. without making
approximations with respect to the distances spanned by these
coordinates. In this paper, we will show that for a special geometry 
of spacetime in which a plane gravitational wave is propagating in a 
flat background, the construction can be carried out to all orders of 
a perturbation expansion leading to analytical formulae for Fermi 
normal coordinates which are valid outside the long-wavelength
regime. Effectively, this amounts to summation of the infinite series 
in the perturbation expansion. We will also consider the closely
related, {\emph{optical coordinates}} that are built upon null 
geodesics instead of spacelike geodesics of the Fermi scheme. Optical 
coordinates represent another choice for a normal-coordinate system 
available to the observer attempting to study the effects of a 
gravitational wave in his vicinity. Analysis of the boundary-value 
problem for geodesics defining the normal coordinates shows that 
there is one more solution. Namely, a special coordinate system can 
be introduced in which spacelike geodesics extend outward from the 
observer synchronously with the incoming gravitational wave, and which
we will  call {\emph{wave-synchronous coordinates.}} We will show that 
these coordinates yield the exact solution of Peres and Ehlers-Kundt, 
which is why there were no higher-order corrections to the metric 
associated with these coordinates. This also explains the difference 
between the coordinate systems in
\cite{Baskaran:2004} and \cite{Rakhmanov:2005}. The first of these
papers involved Fermi normal coordinates whereas the second 
happened to have the wave-synchronuous system.

The presentation of this paper is organized as
follows. Section~\ref{sect:Intro} contains this Introduction. 
In Section~\ref{sect:normalCoords} we give a brief overview
of Riemann and Fermi normal coordinates. 
In Section~\ref{sect:spacetime} we describe two types of solutions 
for a geodesic in spacetime with a plane gravitational wave. 
In Section~\ref{sect:observer} we introduce an ortho-normal tetrad
associated with an inertial observer. 
In Section~\ref{sect:fermiNorm} we give an explicit construction of 
Fermi normal coordinates based on the solution of the boundary-value
problem for geodesics. 
In Section~\ref{sect:series} we derive the infinite series
representation from the exact formulae. 
In Sections~\ref{sect:optical} and \ref{sect:waveSync} we give explicit  
constructions for two other types of normal coordinates: optical
and wave-synchronous. 
In Section~\ref{sect:physics} we compare the different coordinate
systems using the example of an inertial test mass. The Conclusion is 
given in Section~\ref{sect:conclu}. The Appendix contains explicit 
formulae for the Christoffel coefficients and the Riemann tensor.

\section{Overview of Riemann and Fermi normal coordinates}
\label{sect:normalCoords}

We begin with a brief overview of Riemann and Fermi normal coordinates. 
Consider an arbitrary spacetime with coordinates $x^{\mu}$, 
where $\mu = 0, 1, 2, 3$, implying as usual that the first of these 
coordinates is timelike and the rest are spacelike. Let the associated 
metric be $g_{\mu\nu}$. At any given point, $P_0 = \{ x_0^{\mu} \}$, 
the metric tensor can be diagonalized by an orthogonal transformation 
and the resulting diagonal elements can then be scaled to $\pm 1$, 
rendering the metric in the Minkowski form:
\begin{equation}
   \bar{g}_{\mu\nu}(x_0) = \eta_{\mu\nu} ,
\end{equation}
where $\eta_{\mu \nu} = \mathrm{diag}\{-1,1,1,1\}$. In the vicinity of
this reference point, a coordinate system $\bar{x}^{\mu}$ can be
introduced in such a way that the metric remains as close to the
Minkowski form as possible in curved spacetime. In
particular, one can make sure that there are no first-order
corrections to the metric by enforcing the condition:
\begin{equation}
   \bar{g}_{\mu\nu, \alpha}(x_0) = 0 ,
\end{equation}
where comma stands for differentiation with respect to the new
coordinates. This condition is usually
achieved by making the coordinate lines for the
new coordinate system as close to straight lines as they can possibly be  
in curved spacetime, i.e. along geodesics. 
Note that one cannot impose a similar condition on the
second derivatives of the metric unless the curvature of spacetime
vanishes at this point. The resulting coordinates $\bar{x}^{\mu}$  
are known as Riemann normal coordinates. They are
the closest thing to a Cartesian frame that an observer can build in 
curved spacetime in the vicinity of one point and one instance of time. 
In general, the metric contains nonzero second derivatives and its
expansion near the reference point takes the form:
\begin{equation}\label{riemannNorm}
   \bar{g}_{\mu\nu} = \eta_{\mu\nu} - \frac{1}{3} 
   \hat{R}_{\mu\rho\nu\sigma} \, \bar{x}^{\rho} \bar{x}^{\sigma} + \ldots,
\end{equation}
where the dots stand for higher-order terms. 
In what follows the hat above some function will always mean that the
value of this function is taken at the reference point, e.g., 
\begin{equation}
   \hat{R}_{\mu\rho\nu\sigma} \equiv R_{\mu\rho\nu\sigma}(x_0) .
\end{equation}

The Riemann normal coordinates are tied to the reference point
which serves as the origin for this coordinate system. Take another
point, and the whole construction must be repeated yielding a
coordinate system which may not be connected with the first one in any
obvious way. This observation prompted Fermi \cite{Fermi:1922} 
to introduce a different coordinate system -- one which is built 
around a reference curve rather than a point. In Fermi's
construction, one starts with an arbitrary timelike curve and
chooses the parameter along the curve as the first new coordinate
$\bar{x}^0$. The remaining coordinates $\bar{x}^i$, for $i=1,2,3$ 
are built via geodesics which are 
orthogonal to the reference curve and to each other. This approach 
guarantees that the derivatives of the metric with respect to these
remaining coordinates vanish when evaluated on the reference curve.
The derivative with respect to the first coordinate may not be zero. 
If the reference curve itself is geodesic, then all the first
derivatives of the metric vanish and the resulting coordinates
are called Fermi normal. They are the closest thing to a Cartesian
frame that an inertial observer can build in curved spacetime in 
the vicinity of his worldline. The expansion of the metric near the 
reference curve takes the form:
\begin{eqnarray}
  \bar{g}_{00} & = & - 1 - \hat{R}_{0k0l} \, \bar{x}^k \bar{x}^l +
               \ldots , \label{manasseG00} \\
  \bar{g}_{0j} & = & - 
               \frac{2}{3} \hat{R}_{0kjl} \, \bar{x}^k \bar{x}^l +
               \ldots , \label{manasseG0J} \\
  \bar{g}_{ij} & = & \delta_{ij} - 
               \frac{1}{3} \hat{R}_{ikjl} \, \bar{x}^k \bar{x}^l +
               \ldots , \label{manasseGIJ} 
\end{eqnarray}
which no longer has the symmetry between the timelike and spacelike
coordinates present in the Riemann form (\ref{riemannNorm}). 
Equations (\ref{manasseG00})--(\ref{manasseGIJ}) represent the metric
in Fermi normal coordinates to second order in the perturbation
expansion with respect to distance parameters. They were first 
derived\footnote{The definition for the Riemann tensor adopted
in \cite{Manasse:1963} differs from ours by an overall sign.}  
by Manasse and Misner \cite{Manasse:1963}. Significant effort is
needed to go beyond the second-order approximation and generally such 
calculations can be very complicated. The third- and fourth-order 
approximations for Fermi normal coordinates in arbitrary spacetime 
were derived by Li and Ni \cite{Li:1979a, Li:1979b}.

The next important result was obtained by Fortini and Gualdi
\cite{Fortini:1982} who succeeded in deriving the series expansion to
all orders in the distance parameters for spacetime in which a plane 
gravitational wave is propagating in a flat background. The formulae 
of Fortini and Gualdi were later generalized by Marzlin
\cite{Marzlin:1994} for an arbitrary weak-field geometry of spacetime 
and accelerating observers. Marzlin's formulae for the metric in the
case of a non-accelerating observer are 
\begin{eqnarray}
  \bar{g}_{00} & = & - 1 - 2 \sum_{n=0}^{\infty} \frac{n+3}{(n+3)!}
                \hat{R}_{0k0l,m_1 \ldots m_n} \, 
                \bar{x}^k \bar{x}^l \bar{x}^{m_1} \ldots \bar{x}^{m_n}, 
                \label{FortiniG00} \\
  \bar{g}_{0j} & = & - 2 \sum_{n=0}^{\infty} \frac{n+2}{(n+3)!}
                \hat{R}_{0kjl,m_1 \ldots m_n} \, 
                \bar{x}^k \bar{x}^l \bar{x}^{m_1} \ldots \bar{x}^{m_n},
                \label{FortiniG0J} \\
  \bar{g}_{ij} & = & \delta_{ij} - 
                2 \sum_{n=0}^{\infty} \frac{n+1}{(n+3)!}
                \hat{R}_{ikjl,m_1 \ldots m_n} \, 
                \bar{x}^k \bar{x}^l \bar{x}^{m_1} \ldots \bar{x}^{m_n},
                \label{FortiniGIJ} 
\end{eqnarray}
where comma preceding indices $m_1\ldots m_n$ denotes differentiation
with respect to these coordinates, e.g.,
\begin{equation}
   \hat{R}_{\mu k \nu l,m_1 \ldots m_n} \equiv \left. 
      \frac{\partial^n R_{\mu k \nu l}}{\partial x^{m_1} \ldots 
      \partial x^{m_n}} \right|_{x = x_0} . 
\end{equation}
The infinite series, (\ref{FortiniG00})--(\ref{FortiniGIJ}), 
include the second-order metric of Manasse and Misner, and the third- 
and fourth-order metrics of Li and Ni as special cases.

\section{Spacetime with a plane gravitational wave}
\label{sect:spacetime}

\subsection{Metric tensor and fundamental form}

Within the linearized approach, a gravitational wave propagating in
empty space is described by a small perturbation $h_{\mu \nu}$ to the
otherwise flat metric:
\begin{equation}\label{metricLin}
   g_{\mu \nu} = \eta_{\mu \nu} + h_{\mu \nu} .
\end{equation}
The corresponding fundamental form is defined as 
\begin{equation}
   F = g_{\mu\nu} \, \rmd x^{\mu} \, \rmd x^{\nu} ,
\end{equation}
in some coordinate system $x^{\mu}$. Assume that in the absence of 
the gravitational wave, i.e. when $h_{\mu\nu}=0$, three of these
coordinates become the usual Cartesian coordinates,
\begin{equation}
   x^1 \equiv x,  \quad x^2 \equiv y, \quad x^3 \equiv z .
\end{equation}
and the remaining coordinate becomes time 
\begin{equation}
   x^0 \equiv \tau = c t,
\end{equation}
where $c$ is the speed of light in flat spacetime.  
We will use this naming convention even for curved spacetime,
i.e. when $h_{\mu\nu} \neq 0$, keeping in mind that the meaning of these 
coordinates is fundamentally different from that of the Newtonian
world and that the small correction to the metric (\ref{metricLin}) 
changes their interpretation (see Section~\ref{sect:physics}).

A number of components of the symmetric tensor $h_{\mu \nu}$ can be
set to zero by choosing the transverse and traceless gauge 
\cite{Misner:1973}. The remaining components,
\begin{eqnarray}
    h_{xx} = - h_{yy} & \equiv & h_{+}(\tau + z) , \\
    h_{xy} =   h_{yx} & \equiv & h_{\times}(\tau + z) ,
\end{eqnarray}
represent two independent degrees of freedom of the gravitational
wave, commonly known as the $+$ and $\times$ polarizations. The gauge
fixes the orientation of the coordinate system in such a way that the
gravitational wave is propagating in the negative-$z$ direction and its 
transverse polarizations belong to the $xy$ plane. 
In these coordinates, the fundamental form is given by
\begin{eqnarray}
   F & = & - \rmd\tau^2 + \rmd x^2 + \rmd y^2 + \rmd z^2 + 
           \nonumber \\
     &   & h_{+}(\tau+z) \; ( \rmd x^2 - \rmd y^2 ) + 
           2  h_{\times}(\tau+z) \; \rmd x \, \rmd y . \label{metricForm1}
\end{eqnarray}
One can also introduce a natural measure, $\sigma$, such that
\begin{equation}\label{indicatorFdS}
   F = \epsilon \, \rmd \sigma^2 .
\end{equation}
To ensure that $\rmd \sigma$ is real we choose $\epsilon = 1$ or $-1$ 
depending on whether the fundamental form is positive or negative
definite.

We also introduce two auxiliary coordinates,
\begin{equation}
   u = \tau + z    \qquad {\mathrm{and}} \qquad 
   v = \tau - z ,
\end{equation}
which sometimes will be more convenient than $\tau$ and $z$. For
example, the propagation of the plane front of the gravitational wave
is described by $u = \mathrm{const}$. In terms of $u$ and $v$, 
the fundamental form (\ref{metricForm1}) is given by
\begin{equation}\label{metricForm2}
   F = - \rmd u \, \rmd v + \rmd x^2 + \rmd y^2 +
         h_{+}(u) (\rmd x^2 - \rmd y^2) + 
         2 h_{\times}(u) \, \rmd x \, \rmd y . 
\end{equation}
We do not assume any particular form for the functions 
\begin{equation}
    h_a = h_a(u) , 
      \qquad {\mathrm{where}} \qquad a = +, \times.
\end{equation}
They can be completely arbitrary as long as they represent some 
physically possible waveforms.

Within the linearized theory of gravitation we only need to keep track
of terms which are first order in $h_a$. Second and higher order terms 
are neglected. Hence, we will be freely replacing any expression 
containing $h_a$ with its linear (first-order in $h_a$) approximation 
throughout this paper.

\subsection{Geodesic equation}
\label{sect:geodesic}

The construction of normal coordinates relies on the explicit solution
of the geodesic equation. Here we describe the solution for 
geodesics following closely the derivation in \cite{Rakhmanov:2009}. 
Let $x^{\mu}(\sigma)$ be a continuous curve in this spacetime and 
$p^{\mu}(\sigma)$ be a tangent vector for this curve: 
\begin{equation}
   p^{\mu} = \frac{\rmd x^{\mu}}{\rmd \sigma}  
   \qquad {\mathrm{and}} \qquad   
   p_{\mu} =  g_{\mu\nu} \, p^{\nu} .
\end{equation}
Explicit formulae for the covariant components of the tangent vector are
\begin{eqnarray}
   p_v & = & -\frac{1}{2} \frac{\rmd u}{\rmd \sigma} , \label{pvFromU} \\
   p_x & = & \left[ 1 + h_{+}(u) \right] \frac{\rmd x}{\rmd \sigma} + 
             h_{\times}(u) \frac{\rmd y}{\rmd \sigma} , \label{pxFromXY} \\
   p_y & = & \left[ 1 - h_{+}(u) \right] \frac{\rmd y}{\rmd \sigma} + 
             h_{\times}(u) \frac{\rmd x}{\rmd \sigma} , \label{pyFromXY} \\
   p_u & = & -\frac{1}{2} \frac{\rmd v}{\rmd \sigma} . \label{puFromV}
\end{eqnarray}
Assume that $\sigma$ is the natural measure along the curve. Then by virtue
of (\ref{indicatorFdS}) the tangent vector becomes normalized, 
\begin{equation}\label{normPmuPmu}
   p_{\mu} p^{\mu} = \epsilon ,
\end{equation}
where now the indicator $\epsilon$ takes values $-1, 1$, or $0$,
depending on whether the tangent vector is timelike, spacelike, or null.
The normalization condition, written in terms of the covariant
components, is given by
\begin{equation}\label{normPexplicit}
   - 4 p_{u} p_{v} + p_{x}^2 [1 - h_{+}(u)]  
                  + p_{y}^2 [1 + h_{+}(u)] 
   - 2 p_{x} p_{y} \, h_{\times}(u) = \epsilon .
\end{equation}
Next, assume that the curve $x^{\mu}(\sigma)$ is a geodesic. Then the
tangent vector satisfies the equation
\begin{equation}
   \frac{\rmd p_{\alpha}}{\rmd \sigma} = \frac{1}{2} h_{\mu\nu,\alpha} \,
      p^{\mu} p^{\nu} .
\end{equation}
Or, equivalently, 
\begin{equation}\label{eqForPu}
   \frac{\rmd p_u}{\rmd \sigma} = \frac{1}{2}  
      \left( p_x^2 - p_y^2 \right) h'_{+}(u) + p_x p_y h'_{\times}(u) ,
\end{equation}
and
\begin{equation}
   \frac{\rmd p_v}{\rmd \sigma} = 0, \qquad 
   \frac{\rmd p_x}{\rmd \sigma} = 0, \qquad 
   \frac{\rmd p_y}{\rmd \sigma} = 0.
\end{equation}
Therefore, three components of the tangent vector are constant 
along the geodesic:
\begin{eqnarray}
   p_v & = & p_{v0} , \label{eqForPv} \\
   p_x & = & p_{x0} , \label{eqForPx} \\
   p_y & = & p_{y0} . \label{eqForPy}
\end{eqnarray}
The fourth component, $p_u$, can be found from 
the normalization condition (\ref{normPexplicit}):
\begin{equation}\label{puFromNormMain}
   p_{u} = \frac{1}{4 p_{v0}} \left[ - \epsilon + 
           p_{x0}^2 + p_{y0}^2 - (p_{x0}^2 - p_{y0}^2) \, h_{+}(u) -
         2 p_{x0} p_{y0} \, h_{\times}(u) \right] ,
\end{equation}
where for the moment we assumed that $p_{v0} \neq 0$. It turns out that 
this is not always the case. The solutions for the geodesic equation 
which correspond to $p_{v0} \neq 0$ will be called {\emph{main}} or 
{\emph{non-singular}} and the solutions which correspond $p_{v0} = 0$
will be called {\emph{singular}}. Note that the main or non-singular 
solution allows all three types of geodesics: timelike, spacelike, and
null, whereas the singular solution allows only spacelike geodesics. 
The calculations proceed differently for the singular and non-singular
cases.

\subsection{The non-singular solution of the geodesic equation}
\label{sect:geodesic1}

The main or non-singular solution of the geodesic equation takes place
if 
\begin{equation}
   p_{v0} \neq 0 .
\end{equation}
In this case, integration of equation (\ref{pvFromU}) yields the
solution for $u$:
\begin{equation}\label{eqForU(s)}
   u(\sigma) = u_0 - 2 p_{v0} \sigma ,
\end{equation}
where $u_0$ is the initial value for this coordinate. This solution 
alone completely defines the gravitational wave amplitudes on the
geodesic:
\begin{equation}\label{hOnGeod}
   h_a = h_a[u(\sigma)]  .
\end{equation}
Next, inverting (\ref{pxFromXY}) and (\ref{pyFromXY}) to first order in
$h$, we obtain
\begin{eqnarray}
   \frac{\rmd x}{\rmd \sigma} & = & 
      p_{x0} \, \left\{ 1 - h_{+}[u(\sigma)] \right\} - 
      p_{y0} \, h_{\times}[u(\sigma)] , \label{dx/dt}\\
   \frac{\rmd y}{\rmd \sigma} & = & 
      p_{y0} \, \left\{ 1 + h_{+}[u(\sigma)] \right\} - 
      p_{x0} \, h_{\times}[u(\sigma)] . \label{dy/dt}
\end{eqnarray}
Integration of these equations yields the solution for $x$ and $y$:
\begin{eqnarray}
   x(\sigma) & = & x_0 + p_{x0} \sigma \left[ 1 - f_{+}(\sigma) \right] - 
                   p_{y0} \sigma f_{\times}(\sigma),
                   \label{eqForX(s)} \\
   y(\sigma) & = & y_0 + p_{y0} \sigma \left[ 1 + f_{+}(\sigma) \right] - 
                   p_{x0} \sigma f_{\times}(\sigma),
              \label{eqForY(s)} 
\end{eqnarray}
where $x_0$ and $y_0$ are the initial values for these coordinates. 
In the last two equations we introduced the average polarization 
amplitudes of the gravitational wave,
\begin{equation}\label{defGa}
   f_a(\sigma) = \frac{1}{\sigma} \int_0^{\sigma} 
      h_a[u(\sigma')] \, \rmd \sigma' .
\end{equation}
Note that $|f_a| \leq \max |h_a|$ and therefore, $f_a$ is at most the
same order of magnitude as $h_a$.

Consider now equation (\ref{puFromNormMain}). Since $h_a$ are now
fully defined along the geodesic (\ref{hOnGeod}), this equation yields 
$p_{u}$ as a function of $\sigma$:
\begin{eqnarray}
   p_{u}(\sigma) & = & \frac{1}{4 p_{v0}} \left\{ - \epsilon + 
                       p_{x0}^2 + p_{y0}^2 - \right. \nonumber \\
                 &   & \left. (p_{x0}^2 - p_{y0}^2) \, h_{+}[u(\sigma)] -
                       2 p_{x0} p_{y0} \, h_{\times}[u(\sigma)] \right\} .
                       \label{puFromNorm}
\end{eqnarray}
Then the solution for $v$ can be obtained from (\ref{puFromV}):
\begin{equation}\label{eqForV(s)}
   v(\sigma) = v_0 - 2 \int_0^{\sigma} p_u(\sigma') \, \rmd \sigma' ,
\end{equation}
where $v_0$ is the initial value for this coordinate. We will also
need the solution for $z$ and $\tau$:
\begin{eqnarray}
   z(\sigma) & = & z_0 - p_{v0} \sigma + 
      \int_0^{\sigma} p_u(\sigma') \, \rmd \sigma' , \label{eqForZ(s)} \\
   \tau(\sigma) & = & \tau_0 - p_{v0} \sigma - 
      \int_0^{\sigma} p_u(\sigma') \, \rmd \sigma' . \label{eqForT(s)}
\end{eqnarray}
where $z_0$ and $\tau_0$ are the initial values for these
coordinates ($u_0 = \tau_0 + z_0$ and $v_0 = \tau_0 - z_0$).
This concludes the solution for the geodesic equation in the
non-singular case.

The average amplitudes $f_a$ will play an important role in all the
following calculations. Changing variables in (\ref{defGa}), 
we obtain an alternative definition for $f_a$: 
\begin{equation}\label{defGaAlt}
   f_a(u_0, u) = \frac{1}{u - u_0} \int_{u_0}^u  h_a(u') \, \rmd u' ,
\end{equation}
which will sometimes be more convenient than (\ref{defGa}).

\subsection{The singular solution of the geodesic equation}
\label{sect:geodesic2}

The singular solution of the geodesic equation takes place if 
\begin{equation}\label{pV0zero}
   p_{v0} = 0 .
\end{equation}
This condition together with (\ref{pvFromU}) implies that $u$
is constant along the geodesic:
\begin{equation}\label{eqForUU0}
   u(\sigma) = u_0 .
\end{equation}
Since $u$ is constant, the amplitudes of the gravitational wave are
also constant along the geodesic:
\begin{equation}
   h_{a}(u) = h_a(u_0) ,
\end{equation}
and therefore, 
\begin{equation}
   f_a(\sigma) = h_a(u_0) .
\end{equation}
Then the solution for $x$ and $y$ can be obtained by integrating
(\ref{dx/dt}) and (\ref{dy/dt}):
\begin{eqnarray}
   x(\sigma) & = & x_0 + p_{x0} \, \sigma \, [1 - h_{+}(u_0)] - 
                         p_{y0} \, \sigma \, h_{\times}(u_0) ,
                   \label{eqForXinWS} \\
   y(\sigma) & = & y_0 + p_{y0} \, \sigma \, [1 + h_{+}(u_0)] - 
                         p_{x0} \, \sigma \, h_{\times}(u_0) .
                   \label{eqForYinWS}
\end{eqnarray}
Note that we can no longer use the normalization condition 
(\ref{normPexplicit}) to find $p_u$. Instead, we shall find $p_u$
by solving (\ref{eqForPu}). Not only $h_a(u)$ are constant along 
the geodesic but also $h_a'(u)$ are constant. Consequently, the entire
right-hand side of (\ref{eqForPu}) is constant along the
geodesic. Denote this constant by $2A$: 
\begin{equation}\label{geodPuWS}
   2 A \equiv \frac{1}{2} \left( p_{x0}^2 - p_{y0}^2 \right) 
         h'_{+}(u_0) + p_{x0} p_{y0} \, h'_{\times}(u_0) .
\end{equation}
Then equation (\ref{eqForPu}) becomes  
\begin{equation}\label{puWS}
   \frac{\rmd p_u}{\rmd \sigma} = 2 A .
\end{equation}
Its integration yields
\begin{equation}
   p_u(\sigma) = 2 A \, \sigma + B ,
\end{equation}
where $B$ is an arbitrary constant. Finally, integrating 
(\ref{puFromV}), we obtain
\begin{equation}\label{eqForVinWS}
   v(\sigma) = v_0 - 2 A \sigma^2 - 2 B \sigma ,
\end{equation}
where $v_0$ is the initial value for this coordinate. We will also
need the solution for $z$ and $\tau$:
\begin{eqnarray}
   z(\sigma)    & = & z_0    + A \sigma^2 + B \sigma , 
      \label{eqForZinWS} \\
   \tau(\sigma) & = & \tau_0 - A \sigma^2 - B \sigma ,
      \label{eqForTinWS}
\end{eqnarray}
where $z_0$ and $\tau_0$ are the initial values for these
coordinates ($u_0 = \tau_0 + z_0$ and $v_0 = \tau_0 - z_0$). 
This concludes the solution for the geodesic equation in the singular
case.

\section{Inertial observer and the orthonormal tetrad}
\label{sect:observer}

\subsection{The observer's worldline}

One consequence of the solution for the geodesic equation is of
particular interest. This is the notion that an inertial mass which is
initially at rest in the field of a plane gravitational wave remains
at rest indefinitely \cite{Misner:1973, Grishchuk:1977}. Also, an
inertial mass which is moving along the direction of the gravitational 
wave propagation ($\pm z$) remains unaffected by the gravitational
wave. This can be easily seen from the main solution for geodesics
given by (\ref{eqForX(s)}), (\ref{eqForY(s)}), (\ref{eqForZ(s)}), 
and (\ref{eqForT(s)}), in which we set $\epsilon = -1$ and replace 
$\sigma$ with $s$. Note that in this case, the geodesic is
timelike and the affine parameter $s$ stands for the proper time.  
Assume that at the beginning of the geodesic $\rmd x/\rmd s = 0$
and $\rmd y/\rmd s = 0$, which means that $p_{x0} = p_{y0} = 0$. 
Therefore, equations (\ref{eqForX(s)}) and (\ref{eqForY(s)}), become
\begin{equation}\label{atRestXY}
   x(s) = x_0 \qquad {\mathrm{and}} \qquad  y(s) = y_0 .
\end{equation}
Next, equations (\ref{eqForZ(s)}) and (\ref{eqForT(s)}) can be written
as
\begin{eqnarray}
   z(s) & = & z_0 + p^z \, s ,\\
   \tau(s) & = & \tau_0 + p^{\tau} \, s , \label{obsTauS}
\end{eqnarray}
where the constant $p^z$ is defined by $\rmd z/\rmd s$ at the initial
point on the geodesic, and $p^{\tau} = \sqrt{1 + (p^z)^2}$. 
Thus, the mass which was initially moving along the $z$ direction
continues its motion seemingly un-affected by the gravitational wave.

Since the metric in (\ref{metricForm1}) is invariant under Lorentz
transformations in the $z\tau$ plane, we can always transfer to the 
co-moving coordinate system to achieve $p^z = 0$. Then, in addition to
(\ref{atRestXY}), we will have
\begin{equation}
   z(s) = z_0 .
\end{equation}
In these coordinates $p^{\tau} = 1$. Consequently, 
\begin{equation}\label{propTimeObs}
   \tau(s) = s ,
\end{equation}
where we set to zero the arbitrary constant of integration $\tau_0$
introduced in equation (\ref{obsTauS}). Therefore, the proper time of 
the mass, $s$, coincides with the coordinate time, $\tau$, for an
arbitrary gravitational wave $h_a(u)$.

All these arguments can also be applied to an observer who is moving
freely in the field of a gravitational wave, i.e. an inertial observer. 
Namely, the formulae in this Section can be used to describe the
observer's worldline, which is a timelike geodesic with affine
parameter $s$ measuring the observer's proper time. It follows then
that an observer who is initially at rest in the field of a
gravitational wave remains at rest indefinitely. In this regard, he
appears to be un-affected by the gravitational wave. The observer's
clock, which was initially synchronized with the coordinate time,
continues reading the coordinate time even in the presence of the
gravitational wave. Also, the clock appears to be un-affected by the
gravitational wave. These notions turn out to be artifacts of the
present coordinate system (TT gauge). In normal coordinates, an
inertial mass will be moving in response to gravitational waves (see 
Section~\ref{sect:physics}) and the clock will not stay synchronized 
with the coordinate time.

\subsection{Orthonormal tetrad}

To define a quasi-Cartesian coordinate system in his vicinity, an
observer first needs to introduce four basis vectors, a tetrad, which
we denote here as $\lambda^{\bar{\alpha}}$, where $\bar{\alpha}=0,1,2,3$. 
The components of the basis vectors are 
$(\lambda^{\bar{\alpha}})_{\mu} \equiv \lambda^{\bar{\alpha}}{}_{\mu}$. 
Here the index with overline denotes the vector's order number 
whereas the index without overline denotes the vector's component, 
which in this case is covariant. The contravariant components are
defined according to the usual rule:
\begin{equation}
   \lambda^{\bar{\alpha}\mu} = g^{\mu\nu} \,
   \lambda^{\bar{\alpha}}{}_{\nu} . 
\end{equation}
By definition, the basis vectors are ortho-normal:
\begin{equation}
   \lambda^{\bar{\alpha}}{}_{\mu} \, 
   \lambda^{\bar{\beta}}{}^{\mu} = \eta^{\bar{\alpha}\bar{\beta}} .
\end{equation}
Since the observer needs to carry the tetrad with him to the future,
the basis vectors must be transported along the observer's worldline,
which is the reference curve. The result is a one-parameter family: 
$\lambda^{\bar{\alpha}}{}_{\mu}(s)$, where $s$ is the measure along
the reference curve. For a tetrad to be non-rotating, it must be
carried along the curve by means of Fermi-Walker transport 
\cite{Manasse:1963}. If the observer is freely falling, the
Fermi-Walker transport becomes parallel transport: 
\begin{equation}\label{transpLambda}
   \frac{\rmd \lambda^{\bar{\alpha}}{}_{\nu}}{\rmd s} = 
   \Gamma^{\mu}{}_{\beta\nu} \, p^{\beta} \lambda^{\bar{\alpha}}{}_{\mu} ,
\end{equation}
where $\Gamma^{\mu}{}_{\beta\nu}$ are the Christoffel coefficients
(\ref{app:Christoff-Riemann}). Note that these equations are satisfied
independently by each vector $\mathbf{\lambda}^{\bar{\alpha}}$.

As we have seen, an observer at rest remains at rest indefinitely. 
In this case, $p^{\mu} = \{ 1, 0, 0, 0\}$ and equation 
(\ref{transpLambda}) becomes
\begin{equation}\label{transpLambda2}
   \frac{\rmd \lambda^{\bar{\alpha}}{}_{\nu}}{\rmd s} = 
      \Gamma^{\mu}{}_{0\nu} \,
      \lambda^{\bar{\alpha}}{}_{\mu} .
\end{equation}
This equation has infinite number of solutions all of which
correspond to tetrads connected to each other by arbitrary 
rotations, and which, therefore, are all equivalent. To remove this
degeneracy, we assume that in the absence of the gravitational wave, 
the tetrad is consonant with the coordinate system, i.e. 
$\lambda^{\bar{\alpha}}{}_{\mu} = \delta^{\bar{\alpha}}{}_{\mu}$ if $h_a = 0$.
Then the solution of (\ref{transpLambda}) becomes unique and to first order
in $h$ is given by 
\begin{equation}\label{covarLambda}
   \lambda^{\bar{\alpha}}{}_{\mu}(s) = \delta^{\bar{\alpha}}{}_{\mu} + 
   \frac{1}{2} h^{\bar{\alpha}}{}_{\mu}(s) ,
\end{equation}
or, equivalently, 
\begin{equation}\label{contravLambda}
   \lambda^{\bar{\alpha}{\mu}}(s) = \eta^{\bar{\alpha}{\mu}} - 
   \frac{1}{2} h^{\bar{\alpha}{\mu}}(s) .
\end{equation}
Parallel transport of the basis vectors along the geodesic can now be 
achieved simply by advancing parameter $s$ in the right-hand side of 
(\ref{covarLambda}) and (\ref{contravLambda}).

Naturally, the use of this tetrad is limited to freely falling observers. 
To consider an observer on Earth one has to include the observer's 
acceleration due to the normal forces which compensate the 
gravitational pull of the planet or the acceleration due to rotations 
of the laboratory frame (e.g. \cite{Ni:1978, Li:1979b, Marzlin:1994}).

\section{Fermi normal coordinates}
\label{sect:fermiNorm}

We now proceed to the construction of Fermi normal coordinates 
associated with the worldline of an inertial observer. While the 
principal directions for the new coordinate system are set by the 
tetrad, the values for these coordinates are defined by the distances 
along the geodesics which originate from the location of the observer 
and reach every point in the observer's vicinity. Thus, we need to
solve a boundary-value problem: for every point in this spacetime we 
need to find a geodesic which connects it with the observer at a
certain point on his worldline. The choice of this point is not unique
and will lead to different constructions of the quasi-Cartesian frames.
Only one of them corresponds to Fermi normal frame. The geodesic which 
connects the observer with the particular point in spacetime will be 
called the {\emph{connecting geodesic}} so as to distinguish it from the 
{\emph{reference geodesic}}, which is the worldline of the observer. 
(The affine parameter on the connecting geodesic will be denoted by
$\sigma$, whereas the parameter on the reference geodesic will remain
$s$.)

\subsection{Boundary-value problem for the connecting geodesics}
\label{sect:BVproblem}

\begin{figure}[t]
   \centering\includegraphics[width=0.75\textwidth]{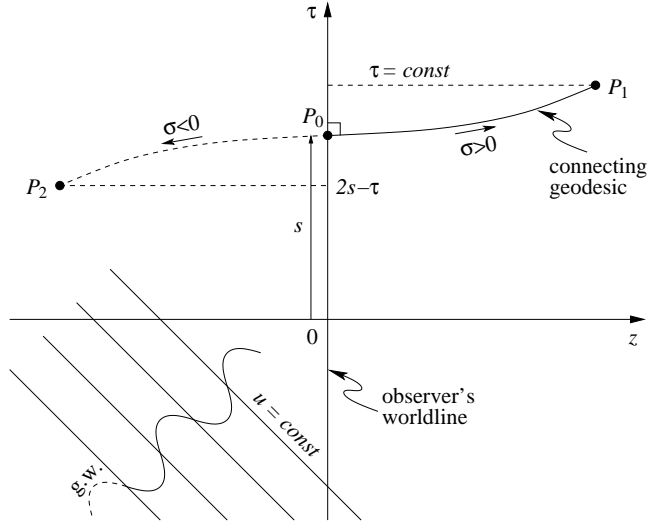}
   \caption{The observer's worldline $z=0$ and the connecting
     geodesic $P_0 P_1$ in the boundary-value problem for Fermi 
     normal coordinates. Negative values of parameter $\sigma$
     correspond to the extension of the connecting geodesic into the
     past. The inverse of $P_1 = \{\tau, x, y, z\}$ is 
     $P_2 \approx \{2s-\tau, -x, -y, -z \}$.}
   \label{fig:fermiNC}
\end{figure}

Take an arbitrary point in spacetime $P_1 = \{ x^{\mu} \}$ and 
connect it with the worldline of the observer via a spacelike
geodesic, $x^{\mu} =  x^{\mu}(\sigma)$, where the affine parameter
$\sigma$ takes values from interval $[0, \sigma_1]$.
Let the tangent vector along the geodesic be 
$p^{\mu} = \rmd x^{\mu}/ \rmd \sigma$. This connecting geodesic will
intersect the observer's worldline at a some point 
$P_0 = \{ x_0^{\mu} \}$. Following Fermi, we require that the
connecting geodesic cut the observer's worldline orthogonally, i.e. 
\begin{equation}\label{FN:cutOrtho}
   \left. p^{\tau} \right|_{\sigma=0} = 0 .
\end{equation}
Physically, this condition implies that the connecting geodesic is the 
closest thing to the instantaneous line in this spacetime, as 
illustrated in figure \ref{fig:fermiNC}.
Mathematically, equation (\ref{FN:cutOrtho}) defines the location of 
point $P_0$ on the observer's worldline.

In what follows, it will be convenient to think of $P_0$ as
the starting point on the connecting geodesic ($\sigma = 0$)  
and $P_1$ as the end point ($\sigma = \sigma_1$). 
Consider the main (non-singular) solution for a geodesic from 
Section \ref{sect:geodesic1} and apply it to the connecting geodesic
$P_0 P_1$. Note that we cannot take the singular solution because 
the constraint (\ref{pV0zero}) is incompatible with the condition 
for orthogonality (\ref{FN:cutOrtho}).

To simplify the following calculations we make a few notational
changes. Assume that the observer is located at the spatial origin 
of the coordinate system:
\begin{equation}\label{xyz0}
   x_0 = y_0 = z_0 = 0 .
\end{equation}
Since the coordinate time at point $P_0$ coincides with the proper
time of the observer (\ref{propTimeObs}), we have 
\begin{equation}
   \tau_0 = s , \qquad {\mathrm{and}} \qquad u_0 = v_0 = s .
\end{equation}
In this notation, the connecting geodesic starts at 
$P_0 = \{ s, 0, 0, 0 \}$ and ends at $P_1 = \{ \tau, x, y, z \}$. 
Alternatively, we can use the $u, v$ coordinates instead of $\tau, z$
and denote the end point as $P_1 = \{ u, x, y, v \}$. Then the
starting point will be $P_0 = \{ s, 0, 0, s \}$. Also we introduce
compact notation for $h_a$ at $P_1$ and $P_0$:
\begin{eqnarray}
   h_a       & \equiv & \left. h_a \right|_{P_1} = h_a(u) ,
       \label{h_at_P1} \\
   \hat{h}_a & \equiv & \left. h_a \right|_{P_0} = h_a(s) . 
       \label{h_at_P0}
\end{eqnarray}

The solution for the geodesic equation (Section~\ref{sect:geodesic1})
connects the coordinates of points $P_0$ and  $P_1$: 
\begin{eqnarray}
   u & = & s - 2 p_{v0} \sigma_1 ,\label{FN:eqU} \\
   x & = & p_{x0} \, \sigma_1 \, (1 - f_{+}) - 
           p_{y0} \, \sigma_1 \, f_{\times} , \label{FN:eqX} \\
   y & = & p_{y0} \, \sigma_1 \, (1 + f_{+}) - 
           p_{x0} \, \sigma_1 \, f_{\times} , \label{FN:eqY} \\
   v & = & s - \frac{\sigma_1}{2 p_{v0}} \left[ - \epsilon +
           p_{x0}^2 + p_{y0}^2 - (p_{x0}^2 - p_{y0}^2) f_{+} - 
           2 \, p_{x0} p_{y0} f_{\times} \right] , \label{FN:eqV}
\end{eqnarray}
where $\epsilon = 1$ and $f_a$ is given by
\begin{equation}\label{FN:Fa}
   f_a = \frac{1}{u - s} \int_s^u  h_a(u') \, \rmd u' .
\end{equation}
The boundary-value problem is to determine arbitrary constants 
$p_{v0}$, $p_{x0}$, $p_{y0}$, $s$, $\sigma_1$ in terms of the
coordinates of point $P_1$.

Due to nonlinear nature of the boundary-value problem the solution
consists of several steps. 
First, we find the components of the tangent vector from 
(\ref{FN:eqU})--(\ref{FN:eqY}):
\begin{eqnarray}
   p_{v0} & = & - \frac{1}{2 \sigma_1}(u - s) ,\label{FN:pv0} \\
   p_{x0} & = & (1 + f_{+}) \frac{x}{\sigma_1} + 
              f_{\times} \, \frac{y}{\sigma_1} , \label{FN:px0} \\
   p_{y0} & = & (1 - f_{+}) \frac{y}{\sigma_1} + 
              f_{\times} \, \frac{x}{\sigma_1} . \label{FN:py0}
\end{eqnarray}
We shall not be concerned with the fact that $s$ and $\sigma_1$ are
unknown at this point. We will have to come back to these equations 
after we determine $s$ and $\sigma_1$. Substituting 
(\ref{FN:pv0})--(\ref{FN:py0}) into (\ref{FN:eqV}), we obtain 
\begin{equation}\label{FN:nonlin1}
   (u - s)(v - s) = - \sigma_1^2 + x^2 + y^2 + 
      (x^2 - y^2) \, f_{+} + 2 x y \, f_{\times} .
\end{equation}
The next step involves the orthogonality condition (\ref{FN:cutOrtho}). 
Since $p^{\tau} = - p_{v0} - p_{u}$ we can express this condition as 
\begin{equation}
   \left. p_{u} \right|_{\sigma=0} = - p_{v0} .
\end{equation}
With the explicit form for $p_{u}$, equation (\ref{puFromNorm}), this 
condition becomes
\begin{equation}\label{FN:ortho}
   4 p_{v0}^2 = 1 - p_{x0}^2 - p_{y0}^2 + 
      (p_{x0}^2 - p_{y0}^2) \hat{h}_{+} + 2 p_{x0} p_{y0} \hat{h}_{\times} .
\end{equation}
Substituting (\ref{FN:pv0})--(\ref{FN:py0}) in (\ref{FN:ortho}), 
we obtain
\begin{equation}\label{FN:nonlin2}
   (u - s)^2 = \sigma_1^2 - x^2 - y^2 - (x^2 - y^2)(2f_{+} - 
      \hat{h}_{+}) - 2 x y (2f_{\times} - \hat{h}_{\times}) .
\end{equation}
The nonlinear equations (\ref{FN:nonlin1}) and (\ref{FN:nonlin2}) 
are the key formulae in the boundary-value problem.

We can now find the solution for $s$. Combining (\ref{FN:nonlin2})
with (\ref{FN:nonlin1}), we eliminate $\sigma_1^2$ with the result 
\begin{equation}\label{FN:eqS2}
   2(s - \tau)(u - s) = (x^2 - y^2)(f_{+} - \hat{h}_{+}) + 
                        2xy (f_{\times} - \hat{h}_{\times}) .
\end{equation}
This equation can be written in the following equivalent form: 
\begin{equation}\label{FN:eqS}
   s = \tau + \frac{1}{2(u - s)} \left[ (x^2 - y^2)(f_{+} - 
       \hat{h}_{+}) + 2xy (f_{\times} - \hat{h}_{\times}) \right] .
\end{equation}
It defines $s$ as an implicit function of the coordinates of the end
point. One can use this equation as an iteration formula to determine 
$s$ because the terms in the square brackets are of order $h$.  
We should not be concerned with the apparent singularity of this
formula as $u \rightarrow s$. One can easily prove that 
\begin{equation}\label{faha_reg}
   \frac{f_a - \hat{h}_a}{u - s} \rightarrow \frac{1}{2} h_a'(s)
\end{equation}
in this limit.

For completeness, we obtain an explicit formula for $s$ to first 
order in $h$. The orthogonality condition (\ref{FN:cutOrtho}) 
implies that in the absence of the gravitational wave $s = \tau$.
Then $u - s \approx z$ to first order in $h$. We can substitute this 
approximation in (\ref{FN:eqS}) and obtain 
\begin{equation}\label{FN:solSfinal}
   s \approx \tau + \frac{1}{2 z} \left[ (x^2 - y^2)(f_{+} - \hat{h}_{+}) + 
       2 x y (f_{\times} - \hat{h}_{\times}) \right] .
\end{equation}
Moreover, to first order in $h$ we can replace every occurrance of $s$ 
with $\tau$ in the right-hand side of this equation. Then 
\begin{eqnarray}
   \hat{h}_a & \approx & h_a(\tau) ,\\
   f_a & \approx & \frac{1}{z} \int_{\tau}^{\tau + z}  h_a(u') \,\rmd u' .
   \label{FN_fa_approx}
\end{eqnarray}
With these approximations, equation (\ref{FN:solSfinal}) yields an 
explicit solution for $s$ and this solution is unique. 
Finally, note that the formulae for $s$ and $f_a$, equations 
(\ref{FN:solSfinal}) and (\ref{FN_fa_approx}), are finite in the 
limit $z \rightarrow 0$. This can be easily seen by substituting the 
Taylor expansion of $h_a(\tau + z)$ in powers of $z$ in
(\ref{FN_fa_approx}).

We can now determine $\sigma_1$ from equation (\ref{FN:nonlin1}). 
Knowing the solution for $s$ and using the fact that $s - \tau$ is of
order $h$, we can approximate (\ref{FN:nonlin1}) as 
\begin{equation}
   \sigma_1^2 = r^2 + (x^2 - y^2) \, f_{+} + 2 x y \, f_{\times} ,
\end{equation}
where $r = \sqrt{x^2 + y^2 + z^2}$. Naturally, there are two solutions:
\begin{equation}
   \sigma_1 = \pm \left[ r^2 + (x^2 - y^2) \, f_{+} + 
            2 x y \, f_{\times} \right]^{1/2} .
\end{equation}
The solution with the ``$-$'' sign corresponds to the extension of the
geodesic beyond the point $P_0$ into the past. The extension 
terminates at point $P_2$ which is the inverse of point $P_1$. 
Therefore, we can safely discard this solution. Taking the ``$+$'' sign 
and keeping only terms first order in $h$, we obtain the final
solution for $\sigma_1$:
\begin{equation}
   \sigma_1 \approx r + \frac{1}{2r} \left[ 
      (x^2 - y^2) \, f_{+} + 2 x y \, f_{\times} \right] .
\end{equation}
Once $s$ and $\sigma_1$ are known, we can return to equations 
(\ref{FN:pv0})--(\ref{FN:py0}) and complete determination 
of $p_{v0}$, $p_{x0}$, $p_{y0}$ by substituting in them the 
explicit formulae for $s$ and $\sigma_1$.
We have thus obtained the solution for the boundary-value problem 
and showed that this solution is unique.

\subsection{Coordinate transformation rules}

With the explicit formulae for the connecting geodesic we can now 
proceed to the construction of Fermi normal coordinates. Let 
$p^{\nu}(\sigma)$ be the tangent vector on the connecting
geodesic. Then the normal coordinates \cite{Synge:1960} of point 
$P_1$ are defined according to 
\begin{equation}\label{defNormCoord}
   \bar{x}^{\mu} \equiv x_0^{\mu} + \lambda^{\bar{\mu}}{}_{\nu} \left. 
      p^{\nu} \right|_{\sigma=0} \sigma_1 ,
\end{equation}
or, equivalently,
\begin{equation}\label{defNormCoord2}
   \bar{x}^{\mu} \equiv x_0^{\mu} + \lambda^{\bar{\mu}{\nu}} \left. 
      p_{\nu} \right|_{\sigma=0} \sigma_1 .
\end{equation}
Note that $p_{v0}, p_{x0}, p_{y0}$ are constant along the geodesic and
for these components we can omit the sign $\left. \right|_{\sigma=0}$ .

Consider first the transformation of $\tau$. By virtue of the orthogonality 
condition (\ref{FN:cutOrtho}), we have 
\begin{equation}
   \bar{\tau} \equiv s + \left. p^{\tau} \right|_{\sigma=0} \sigma_1 = s .
\end{equation}
The transformation of the $z$ coordinate can be found as follows:
\begin{eqnarray}
   \bar{z} & \equiv & \lambda^{\bar{z} z} \left.
                      p_z \right|_{\sigma=0} \sigma_1 \nonumber \\
           & = & - 2 p_{v0} \sigma_1 \nonumber \\
           & = & u - s ,
\end{eqnarray}
where we used the fact that 
$ \left. p_z \right|_{\sigma=0} = - 2 p_{v0}$ which follows from 
the orthogonality condition (\ref{FN:cutOrtho}). Thus, the first two 
equations for the new coordinates are
\begin{eqnarray}
   \bar{\tau} & = & s ,     \label{FNbarT} \\
   \bar{z}    & = & u - s . \label{FNbarZ}
\end{eqnarray}
Here the complexity of the coordinate transformation is hidden in $s$ 
which is a function of $\tau, x, y, z$, given by (\ref{FN:eqS}) or
(\ref{FN:solSfinal}). From equations (\ref{FNbarT}) and (\ref{FNbarZ}) 
we can see that
\begin{eqnarray}
  \bar{u} & = & u ,\label{FNeqForU} \\
   \bar{v} & = & v + \frac{1}{u - s} \left[ 
             (x^2 - y^2)(f_{+} - \hat{h}_{+}) + 
             2 x y  (f_{\times} - \hat{h}_{\times}) \right] .\label{FNeqForV}
\end{eqnarray}

Consider now the transformation of the $x$ coordinate. From 
definition (\ref{defNormCoord2}) and also (\ref{xyz0}) we find 
\begin{eqnarray}
   \bar{x} & \equiv & 
          \lambda^{\bar{x} x} p_{x0} \sigma_1 + 
          \lambda^{\bar{x} y} p_{y0} \sigma_1 \nonumber \\
    & = & \left(1 -\frac{1}{2} \hat{h}_{+}     \right) p_{x0} \sigma_1 + 
          \left(  -\frac{1}{2} \hat{h}_{\times} \right) p_{y0} \sigma_1 .
\end{eqnarray}
By substituting the formulae for $p_{x0}$ and $p_{y0}$ from  
(\ref{FN:px0}) and (\ref{FN:py0}), and keeping the terms first order
in $h$ only, we obtain
\begin{equation}\label{FNeqForX}
   \bar{x} = x + x \left( f_{+} - \frac{1}{2} \hat{h}_{+} \right) + 
       y \left( f_{\times} - \frac{1}{2} \hat{h}_{\times} \right) .
\end{equation}
Similar steps lead to the transformation rule for the $y$ coordinate:
\begin{equation}\label{FNeqForY}
   \bar{y} = y - y \left( f_{+} - \frac{1}{2} \hat{h}_{+} \right) + 
       x \left( f_{\times} - \frac{1}{2} \hat{h}_{\times} \right) .
\end{equation}
Finally, we rewrite (\ref{FNbarZ}) and (\ref{FNbarT}), replacing $s$
with its nonlinear representation (\ref{FN:eqS})
\begin{eqnarray}
   \bar{z} & = & z - \frac{1}{2(u - s)} \left[ 
       (x^2 - y^2)(f_{+} - \hat{h}_{+}) +
       2 x y (f_{\times} - \hat{h}_{\times}) \right] , 
       \label{FNeqForZ} \\
   \bar{\tau} & = & \tau + \frac{1}{2(u - s)} \left[ 
       (x^2 - y^2)(f_{+} - \hat{h}_{+}) +
       2 x y (f_{\times} - \hat{h}_{\times}) \right] . 
       \label{FNeqForT}
\end{eqnarray}
We have thus obtained the formulae for Fermi normal coordinates
$\bar{x}, \bar{y}, \bar{z}, \bar{\tau}$ of point $P_1$ in terms of
its TT coordinates $x,y,z,\tau$.

Even though the term $(u-s)$ appears in the denominators of 
(\ref{FNeqForZ}) and (\ref{FNeqForT}), the corresponding fractions  
are not divergent. We have already seen in (\ref{faha_reg}) 
that the function,  
\begin{equation}\label{defHfunct}
   H_a \equiv \frac{f_a - \hat{h}_{a} }{u - s} 
            = \frac{1}{(u - s)^2} 
              \int_s^u \left[ h_a(u') - h_a(s) \right] \rmd u' , 
\end{equation}
has a finite limit for $u \rightarrow s$. Using this function, we can
present the coordinate transformation formulae in an explicitly regular
form:
\begin{eqnarray}
   \bar{x} & = &  x + \frac{1}{2} x \, \hat{h}_{+} +
                      \frac{1}{2} y \, \hat{h}_{\times} + 
                  (u - s) \left( x H_{+} + y H_{\times} \right) , 
                  \label{FNequivX} \\
   \bar{y} & = &  y - \frac{1}{2} y \, \hat{h}_{+} +
                      \frac{1}{2} x \, \hat{h}_{\times} - 
                  (u - s) \left( y H_{+} - x H_{\times} \right) ,\\
   \bar{z} & = & z    - \frac{1}{2} (x^2 - y^2) H_{+} - xy H_{\times} ,\\
\bar{\tau} & = & \tau + \frac{1}{2} (x^2 - y^2) H_{+} + xy H_{\times} .
                 \label{FNequivT}
\end{eqnarray}
This form will be particularly useful for series expansions.

\subsection{Metric in Fermi normal coordinates}
\label{sect:indMetric}

To obtain the metric in Fermi normal coordinates, we need to
invert the coordinate transformation rules 
(\ref{FNequivX})--(\ref{FNequivT}). 
To first order in $h$, the inverse formulae can be written as
\begin{eqnarray}
   x & = & \bar{x} -   \frac{1}{2} \bar{x} \, \hat{h}_{+} -
                       \frac{1}{2} \bar{y} \, \hat{h}_{\times} - 
               \bar{z} \left( \bar{x} H_{+} + \bar{y} H_{\times} \right) , 
                       \label{invFNforX} \\
   y & = & \bar{y} +   \frac{1}{2} \bar{y} \, \hat{h}_{+} -
                       \frac{1}{2} \bar{x} \, \hat{h}_{\times} + 
               \bar{z} \left( \bar{y} H_{+} - \bar{x} H_{\times} \right) , 
                       \label{invFNforY} \\
   z & = & \bar{z}   + \frac{1}{2} (\bar{x}^2 - \bar{y}^2) H_{+} + 
                                    \bar{x} \bar{y} H_{\times} ,
                       \label{invFNforZ} \\
   \tau & = & \bar{\tau} - \frac{1}{2} (\bar{x}^2 - \bar{y}^2) H_{+} - 
                                        \bar{x} \bar{y} H_{\times} ,
                       \label{invFNforT}
\end{eqnarray}
where we replaced $u-s$ with $\bar{z}$. Then $H_a$ is given by 
\begin{equation}\label{defHfunct2}
   H_a = \frac{1}{\bar{z}} \left( f_a - \hat{h}_a \right) ,
\end{equation}
in which $\hat{h}_a$ and $f_a$ must be viewed as functions of the 
new coordinates:
\begin{eqnarray}
   \hat{h}_a & = & h_a(\bar{\tau}),\\
   f_a & = & \frac{1}{\bar{z}} \int_{\bar{\tau}}^{\bar{\tau} + \bar{z}} 
             h_a(u') \, \rmd u' . \label{faNewCoord}
\end{eqnarray}
We then substitute the inverse transformation rules into the
fundamental form (\ref{metricForm1}) and group together all terms 
containing the same binomial $\rmd \bar{x}^{\mu} \rmd \bar{x}^{\nu}$. 
(At this step, it would be simpler to use $u,v$ coordinates instead 
of $z,\tau$.) The resulting formulae for the metric,
\begin{equation}
   \bar{g}_{\mu\nu} = \eta_{\mu\nu} + C_{\mu\nu} ,
\end{equation}
are somewhat complicated. However, they can be greatly simplified if 
we introduce the following functions:
\begin{eqnarray}
   P_a(s, u) & = &  h_a(u) + h_a(s) - 
              \frac{2}{u - s} \int_s^u  h_a(u') \, \rmd u' ,
              \label{defPAlong}\\
   Q_a(s, u) & = & h_a(u) - \frac{1}{2} (u - s) \, h'_a(s) - 
              \frac{1}{u - s} \int_s^u  h_a(u') \, \rmd u' ,
              \label{defQAlong}
\end{eqnarray}
which we will write compactly as 
\begin{eqnarray}
   P_a & = & h_a + \hat{h}_a - 2 f_a , \label{defPA}\\
   Q_a & = & h_a - \frac{1}{2} \bar{z} \, \hat{h}'_a  - f_a .\label{defQA}
\end{eqnarray}
Then the components of the metric can be written as
\begin{eqnarray}
  C_{xx}      & = & P_{+} , \label{FNmetricGxx} \\
  C_{yy}      & = & - P_{+} , \\
  C_{xy}      & = & P_{\times} , \\
  C_{xz}      & = & - \frac{1}{\bar{z}} \left( 
                     \bar{x} P_{+} + \bar{y} P_{\times} \right), \\
  C_{yz}      & = & - \frac{1}{\bar{z}} \left( 
                     \bar{x} P_{\times} - \bar{y} P_{+} \right), \\
  C_{zz}      & = & \frac{1}{\bar{z}^2} \left[ 
                    (\bar{x}^2 - \bar{y}^2) P_{+} + 
                    2 \bar{x} \bar{y} P_{\times} \right] ,\\
  C_{\tau x}   & = & - \frac{1}{\bar{z}} \left( 
                    \bar{x} Q_{+} + \bar{y} Q_{\times} \right),
                    \label{FNmetricGtx} \\
  C_{\tau y}   & = & - \frac{1}{\bar{z}} \left( 
                    \bar{x} Q_{\times} - \bar{y} Q_{+} \right),
                    \label{FNmetricGty} \\
  C_{\tau z}   & = & \frac{1}{\bar{z}^2} \left[  
                    (\bar{x}^2 - \bar{y}^2) Q_{+} + 
                    2 \bar{x} \bar{y} Q_{\times} \right],
                    \label{FNmetricGtz} \\
  C_{\tau\tau} & = & \frac{1}{\bar{z}^2} \left[ 
                    (\bar{x}^2 - \bar{y}^2) (2 Q_{+} - P_{+}) + 
                    2 \bar{x} \bar{y} (2 Q_{\times} - P_{\times}) \right].
                    \label{FNmetricGtt}
\end{eqnarray}
We have thus obtained the metric in Fermi normal coordinates. Since
all derivations here were done to first order in $h$, the formulae 
for the Fermi normal metric are valid as long as 
$|C_{\mu\nu}| \ll 1$. This condition naturally limits the transverse 
coordinates $\bar{x}$ and $\bar{y}$. However, there are no
limitations in the longitudinal direction: the $\bar{z}$ coordinate 
can be completely arbitrary, including the limit 
$\bar{z} \rightarrow 0$. Despite the apparent singularity of some of
the metric components as $\bar{z} \rightarrow 0$, all such expressions
are finite in this limit, as will be shown in Section~\ref{sect:series},

\section{Series expansions in distance parameters}
\label{sect:series}

Fermi normal coordinates and the induced metric have always been 
presented in terms of perturbation series in powers of distance
parameters. In this Section we briefly describe such series expansions
and show how they can be derived from the exact formulae. 
It is within the perturbation series approach that one encounters the
idea of the long-wavelength regime.

\subsection{Infinite series representation}

Consider the Taylor series for $h_a(u)$ defined on the connecting geodesic:
\begin{equation}\label{expandH}
   h_a(u) = \sum_{n=0}^{\infty} \frac{1}{n!} \, 
           (u - s)^n \, h^{(n)}_a(s) .
\end{equation}
The right-hand side of this equation can be viewed as containing only 
Fermi normal coordinates. Indeed, $s = \bar{\tau}$ according to
(\ref{FNbarT}) and $u - s = \bar{z}$ according to
(\ref{FNbarZ}). Therefore, we can write this series as
\begin{equation}\label{expandH2}
   h_a = \sum_{n=0}^{\infty} \frac{{\bar{z}}^n}{n!} \, \hat{h}^{(n)}_a ,
\end{equation}
where for simplicity we suppressed the arguments of the functions 
according to definitions (\ref{h_at_P1}) and (\ref{h_at_P0}). Next, 
we obtain the series expansions for $f_a$ and $H_a$:
\begin{eqnarray}
   f_a & = & \sum_{n=0}^{\infty} \frac{{\bar{z}}^n}{(n + 1)!} \, 
             \hat{h}^{(n)}_a , \label{expandFa}\\
   H_a & = & \sum_{n=1}^{\infty} \frac{{\bar{z}}^{n-1}}{(n + 1)!} \, 
              \hat{h}^{(n)}_a .\label{expandHa}
\end{eqnarray}
Then the series representation for the coordinate transformations
(\ref{invFNforX})--(\ref{invFNforT}) can be obtained by substituting
in them the series for $H_a$ from (\ref{expandHa}). We can also obtain 
the series representation for the coordinate transformations 
(\ref{FNequivX})--(\ref{FNequivT}). In this case, we will need to make 
the following approximations: $\bar{z} \approx z$ and $s \approx \tau$ 
to make sure that the resulting formulae contain only the TT
coordinates in their right-hand sides.

The series representation for the induced metric can be found in two 
different ways. One can calculate the metric from the series
representation for the coordinate transformations. Or, one can
obtain the metric from the exact formulae in 
Section~\ref{sect:indMetric}. The second method is simpler. Indeed, 
using the Taylor series (\ref{expandH2}) and
(\ref{expandFa}), we find the series expansion for $P_a$ and $Q_a$: 
\begin{eqnarray}
   P_a & = & \sum_{n=2}^{\infty} \frac{n-1}{(n+1)!} 
             \bar{z}^{n} \hat{h}_a^{(n)}, \\ 
   Q_a & = & \sum_{n=2}^{\infty} \frac{n}{(n+1)!}
             \bar{z}^{n} \hat{h}_a^{(n)}.
\end{eqnarray}
Substituting these formulae in 
(\ref{FNmetricGxx})--(\ref{FNmetricGtt}), we obtain the series 
representation for the metric in Fermi normal coordinates. 
These formulae will be written in terms of the derivatives of 
$\hat{h}_a$. We can also write them in terms of the components of 
the Riemann tensor: 
\begin{eqnarray}
  \bar{g}_{00} & = & - 1 - 2 \sum_{n=2}^{\infty} \frac{n+1}{(n+1)!}
                \hat{R}_{0k0l}^{(n-2)} \, 
                \bar{x}^k \bar{x}^l \bar{z}^{n-2}, 
                \label{seriesG00} \\
  \bar{g}_{0j} & = & - 2 \sum_{n=2}^{\infty} \frac{n}{(n+1)!}
                \hat{R}_{0kjl}^{(n-2)} \, 
                \bar{x}^k \bar{x}^l \bar{z}^{n-2},
                \label{seriesG0J} \\
  \bar{g}_{ij} & = & \delta_{ij} - 
                2 \sum_{n=2}^{\infty} \frac{n-1}{(n+1)!}
                \hat{R}_{ikjl}^{(n-2)} \, 
                \bar{x}^k \bar{x}^l \bar{z}^{n-2},
                \label{seriesGIJ} 
\end{eqnarray}
where the summation over the transverse tensor indices $k,l$ is 
implicit, and 
\begin{equation}
   \hat{R}_{\mu \nu \alpha \beta}^{(n)} = \left[
   \frac{d^n}{du^n} R_{\mu \nu \alpha \beta}(u) \right]_{u=s} .
\end{equation}
We have thus reproduced the results of Fortini and Gualdi
\cite{Fortini:1982}. Note also that equations 
(\ref{seriesG00})--(\ref{seriesGIJ}) are a special case of 
formulae (\ref{FortiniG00})--(\ref{FortiniGIJ}) derived by 
Marzlin \cite{Marzlin:1994}.

The series expansions show that there are no singularities in the
coordinate transformation rules and the induced metric in the limit of 
$u \rightarrow s$ or $\bar{z} \rightarrow 0$. Also, we can see that  
$\bar{g}_{\mu\nu} \rightarrow \eta_{\mu\nu}$ in the limit when the
distance parameters vanish.

\subsection{Lowest-order approximations}

Truncation of the infinite series will lead to approximate formulae
for Fermi normal coordinates and the induced metric. For example, the 
lowest order approximation is obtained by truncating the Taylor 
series (\ref{expandH2}) at the first order ($n=1$). At this order, 
the coordinate transformation rules are approximated by 
\begin{eqnarray}
   \bar{x} & \approx & x + 
                       \frac{1}{2} x \, \hat{h}_{+} + 
                       \frac{1}{2} y \, \hat{h}_{\times} +
              \frac{1}{2} \bar{z} \left( x \, \hat{h}'_{+} + 
                                   y \, \hat{h}'_{\times} \right),
                                   \label{1orderX} \\
   \bar{y} & \approx & y - 
                       \frac{1}{2} y \, \hat{h}_{+} + 
                       \frac{1}{2} x \, \hat{h}_{\times} -
              \frac{1}{2} \bar{z} \left( y \, \hat{h}'_{+} - 
                                   x \, \hat{h}'_{\times} \right), \\ 
   \bar{z} & \approx & z - 
                       \frac{1}{4} (x^2 - y^2) \, \hat{h}'_{+} - 
                               \frac{1}{2} x y \, \hat{h}'_{\times}, \\
   \bar{\tau} & \approx & \tau + 
                       \frac{1}{4} (x^2 - y^2) \, \hat{h}'_{+} + 
                               \frac{1}{2} x y \, \hat{h}'_{\times} ,
                                   \label{1orderT}
\end{eqnarray}
where we can replace $\bar{z}$ with $z$ in the right-hand side of
these equations. The induced metric at this order is trivial:  
$\bar{g}_{\mu\nu} \approx \eta_{\mu\nu}$ because non-zero 
corrections to the metric appear only in the second order. 
We have thus reproduced the first-order approximation for 
Fermi normal coordinates which appeared in the early papers of 
Grishchuk \cite{Grishchuk:1977, Grishchuk:1980}. These 
transformation rules were later revisited by Baskaran and Grishchuk in 
their analysis of the response of laser gravitational-wave 
detectors \cite{Baskaran:2004}.

The second-order approximation leads to the well-known formulae of
Manasse and Misner. This can be easily seen from equations 
(\ref{seriesG00})--(\ref{seriesGIJ}) if we truncate the series 
with $n = 2$ terms.  
Another approach would be to take the exact formulae for the metric 
(\ref{FNmetricGxx})--(\ref{FNmetricGtt}) and substitute in them the
second-order approximations:
\begin{eqnarray}
   P_a & \approx & \frac{1}{6} \bar{z}^2 \, \hat{h}''_a , \\
   Q_a & \approx & \frac{1}{3} \bar{z}^2 \, \hat{h}''_a . 
\end{eqnarray}
The result is 
\begin{eqnarray}
  C_{xx} & \approx & \frac{1}{6} \bar{z}^2 \, \hat{h}''_{+} ,
              \label{2orderGxx} \\
  C_{yy} & \approx & - \frac{1}{6} \bar{z}^2 \, \hat{h}''_{+} ,\\
  C_{xy} & \approx & \frac{1}{6} \bar{z}^2 \, \hat{h}''_{\times} ,\\
  C_{xz} & \approx & - \frac{1}{6} \bar{z} \left( \bar{x} \, 
              \hat{h}''_{+} + \bar{y} \, \hat{h}''_{\times} \right) ,\\
  C_{yz} & \approx & - \frac{1}{6} \bar{z} \left( \bar{x} \, 
              \hat{h}''_{\times} - \bar{y} \, \hat{h}''_{+} \right) ,\\
  C_{zz} & \approx & \frac{1}{6} (\bar{x}^2 - \bar{y}^2) 
              \hat{h}''_{+} + \frac{1}{3} \bar{x} \, \bar{y} \, 
              \hat{h}''_{\times} ,\\
  C_{\tau x} & \approx & - \frac{1}{3} \bar{z} \left( \bar{x} \, 
              \hat{h}''_{+} + \bar{y} \, \hat{h}''_{\times} \right) ,
              \label{2orderGtx}\\
  C_{\tau y} & \approx & - \frac{1}{3} \bar{z} \left( \bar{x} \, 
              \hat{h}''_{\times} - \bar{y} \, \hat{h}''_{+} \right) ,
              \label{2orderGty}\\
  C_{\tau z} & \approx & \frac{1}{3} (\bar{x}^2 - \bar{y}^2) 
              \hat{h}''_{+} + \frac{2}{3} \bar{x} \, \bar{y} \, 
              \hat{h}''_{\times} ,\label{2orderGtz} \\
  C_{\tau \tau} & \approx & \frac{1}{2} (\bar{x}^2 - \bar{y}^2) 
              \hat{h}''_{+} + \bar{x} \, \bar{y} \, \hat{h}''_{\times} ,
              \label{2orderGtt}
\end{eqnarray}
which is the explicit form for the Manasse-Misner formulae.

By truncating the Taylor series for $h_a$ one can obtain the approximate 
formulae for Fermi normal coordinates and the induced metric at any 
desired order $n$. In any such approximation, it is assumed that the 
$(n+1)$-order terms are much less than the terms of order $n$.  
For a sinusoidal gravitational wave with wavelength $\lambda$, the 
expansion will be in powers of $\bar{z}/\lambda$. Then the 
higher-order terms become negligible if $|\bar{z}| \ll \lambda$. This 
condition is commonly known as the long-wavelength regime.

\section{Optical coordinates}
\label{sect:optical}

It was pointed out by Synge that the condition for orthogonality of
the connecting geodesic and the observer's worldline is somewhat 
artificial \cite{Synge:1960}. A more natural approach would be to
use a null geodesic for the connecting curve.  
We therefore consider here a different boundary-value problem in which we
replace the spacelike connecting geodesic with a null geodesic. 
The resulting normal coordinates will be called 
optical coordinates -- the name suggested by Synge in his 
analysis of normal coordinates for an arbitrary geometry of spacetime 
\cite{Synge:1960}.

The equations for the boundary-value problem in this case are the same as
those in Section \ref{sect:BVproblem}, namely  
(\ref{FN:eqU})--(\ref{FN:eqV}) with $\epsilon = 0$. 
The only difference is that the condition for orthogonality 
(\ref{FN:cutOrtho}) is replaced with the null condition:
\begin{equation}\label{OC:null}
   p_{\mu} p^{\mu} = 0 .
\end{equation}
Unlike the case of the FN coordinate construction, we will not be able
to determine parameter $\sigma_1$ uniquely from the boundaries. This
is because equations (\ref{FN:eqU})--(\ref{FN:eqV}) become scale
invariant for $\epsilon = 0$, i.e. they are invariant under the 
transformation: $\sigma \rightarrow \kappa \sigma$ and 
$p_{\mu} \rightarrow p_{\mu}/\kappa$ for an arbitrary constant $\kappa$. 
By taking advantage of this scale invariance, we can set $\sigma_1 = 1$.

The null condition (\ref{OC:null}) gives us a formula for $p_u$: 
\begin{equation}\label{OC:eqQu}
   p_u = \frac{1}{4 p_{v0}} \left[ p_{x0}^2 + p_{y0}^2 - 
        (p_{x0}^2 - p_{y0}^2) h_{+} - 2 p_{x0} p_{y0} h_{\times} \right] .
\end{equation}
The solution for the geodesic equation (Section~\ref{sect:geodesic1})
connects the coordinates of points $P_0$ and $P_1$: 
\begin{eqnarray}
   u & = & s - 2 p_{v0} , \label{OC:eqU} \\
   x & = & p_{x0} \left( 1 - f_{+} \right) - p_{y0} f_{\times}, \label{OC:eqX} \\
   y & = & p_{y0} \left( 1 + f_{+} \right) - p_{x0} f_{\times}, \label{OC:eqY} \\
   v & = & s - \frac{1}{2 p_{v0}} \left[ 
                   p_{x0}^2 + p_{y0}^2 - (p_{x0}^2 - p_{y0}^2) f_{+} - 
                   2 \, p_{x0} p_{y0} f_{\times} \right] , \label{OC:eqV} 
\end{eqnarray}
where $f_a$ is given by 
\begin{equation}\label{OC:Fa}
   f_a = \frac{1}{u - s} \int_s^u  h_a(u') \, \rmd u' .
\end{equation}
The boundary-value problem is to determine arbitrary constants 
$p_{v0}$, $p_{x0}$, $p_{y0}$, $s$ in terms of the
coordinates of point $P_1$.

As with FN coordinates, our first step is to express constants
$p_{v0},  p_{x0}, p_{y0}$ from equations (\ref{OC:eqU})--(\ref{OC:eqY}):
\begin{eqnarray}
   p_{v0} & = & - (u - s)/2 ,  \label{OC:qv0} \\
   p_{x0} & = & x \, (1 + f_{+}) + y \, f_{\times} , \label{OC:qx0} \\
   p_{y0} & = & y \, (1 - f_{+}) + x \, f_{\times} , \label{OC:qy0}
\end{eqnarray}
and substitute for them in equation (\ref{OC:eqV}). The result is
\begin{equation}\label{OC:nonlinS}
   (s - \tau)^2 = r^2 + (x^2 - y^2) f_{+} + 2 \, x y f_{\times} .
\end{equation}
This equation contains $s$ in the left-hand side and also in the 
right-hand side as an argument of $f_a$. 
However, $f_a$ are first order in $h$ and therefore this equation can
be solved perturbatively.

\begin{figure}[t]
   \centering\includegraphics[width=0.75\textwidth]{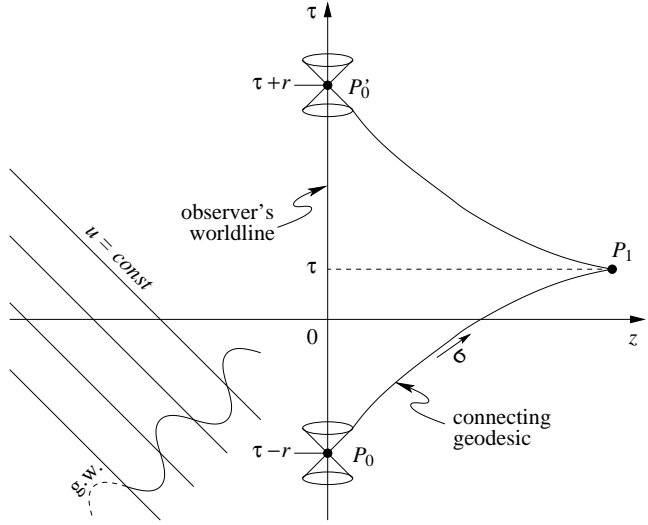}
   \caption{The observer's worldline $z=0$ and the connecting
     geodesic $P_0 P_1$ in the boundary-value problem for 
     optical coordinates. $P_1$ is in the causal future of 
     $P_0 = \{ \tau - r, 0, 0, 0 \}$ and is in the causal past of 
    $P'_0 = \{ \tau + r, 0, 0, 0 \}$.}
   \label{fig:optical}
\end{figure}

In the absence of a gravitational wave, equation (\ref{OC:nonlinS}) 
reduces to the formula for the light cone: $(s - \tau)^2 = r^2$, 
which has two solutions: $s = \tau \pm r$. We choose to have point 
$P_1$ in the causal future of the observer, as shown in figure
\ref{fig:optical}. 
In other words, the connecting geodesic represents the photon 
traveling from $P_0$ to $P_1$. Thus, we take for the 
unperturbed solution $s = \tau - r$. Then to first order in $h$ the
solution of (\ref{OC:nonlinS}) is 
\begin{equation}
   s = \tau -  r - \frac{1}{2r} \left[ (x^2 - y^2) f_{+} + 
       2 \, x y f_{\times} \right] , 
\end{equation}
where now 
\begin{equation}\label{faOPTcoord}
   f_a = \frac{1}{z + r} \int_{\tau - r}^{\tau + z}  h_a(u') \, \rmd u' ,
\end{equation}
which is obtained from (\ref{OC:Fa}) by replacing $s$ with
its zeroth-order approximation $\tau - r$. 
We have thus obtained the solution for $s$ in the boundary-value
problem for the connecting null geodesic and showed
that this solution is unique. We can now return to equations 
(\ref{OC:qv0})--(\ref{OC:qy0}) and complete the calculation of 
$p_{v0}$, $p_{x0}$, $p_{y0}$.

Note that one can also take the second unperturbed solution  
$s = \tau + r$, which corresponds to the situation when point $P_1$ is in
the causal past of the observer. In this case the connecting geodesic
cuts the worldline of the observer at a different point ($P'_0$), 
as shown in figure \ref{fig:optical}. The photon traveling along this
null geodesic will be moving into the past. This choice would lead to 
a slightly different solution for the boundary-value problem and
consequently to a slightly different set of normal coordinates. 
In either case, however, only one null geodesic connects the observer 
with a given point in this spacetime.

\subsection{Coordinate transformation rules}

The coordinate transformations are found from the general rule, 
equation (\ref{defNormCoord2}), which we write here as 
\begin{equation}
   \bar{x}^{\mu} \equiv  x_0^{\mu} + \lambda^{\bar{\mu} {\nu}} 
      \left. p_{\nu} \right|_{\sigma=0}.
\end{equation}
Consider first the transformation of the $u$ coordinate. 
From equation (\ref{OC:qv0}), we obtain 
\begin{equation}
   \bar{u} = s + \lambda^{\bar{u} v} p_{v0} = s + (-2) [ -(u - s)/2 ] ,
\end{equation}
and therefore,
\begin{equation}\label{OC:barU}
   \bar{u} = u .
\end{equation}
Next, consider the transformation of $x$:
\begin{equation}
   \bar{x} = \lambda^{\bar{x} x} p_{x0} + 
             \lambda^{\bar{x} y} p_{y0} . 
\end{equation}
Substituting for $\lambda^{\bar{\mu} \nu}$ and $p_{\nu}$, and keeping 
terms first order in $h$, we obtain
\begin{equation}\label{OC:barX}
  \bar{x} = x + x \left( f_{+} - \frac{1}{2} \hat{h}_{+} \right) + 
                y \left( f_{\times} - \frac{1}{2} \hat{h}_{\times} \right) .
\end{equation}
Similarly, we obtain the transformation of $y$:
\begin{equation}\label{OC:barY}
  \bar{y} = y - y \left( f_{+} - \frac{1}{2} \hat{h}_{+} \right) + 
                x \left( f_{\times} - \frac{1}{2} \hat{h}_{\times} \right) .
\end{equation}
Finally, we consider the transformation of the $v$ coordinate,
\begin{equation}
   \bar{v} \equiv s + \lambda^{\bar{v} u} \left. p_{u} \right|_{\sigma = 0} .
\end{equation}
Using (\ref{OC:eqQu}) and (\ref{OC:qv0})--(\ref{OC:qy0}), 
we obtain
\begin{equation}\label{OC:barV}
   \bar{v} = v + \frac{1}{u - s} \left[ 
             (x^2 - y^2)(f_{+} - \hat{h}_{+}) - 
             2 \, x y  (f_{\times} - \hat{h}_{\times}) \right] .
\end{equation}
It is interesting to note that these transformation rules are formally 
equivalent to those we derived for Fermi normal coordinates, namely  
(\ref{FNeqForU}), (\ref{FNeqForV}), (\ref{FNeqForX}), (\ref{FNeqForY}). 
The only difference is in the definition of the parameter $s$. 
The fact that the formulae for optical coordinates are identical to
those of Fermi normal coordinates is not related to the symmetries of
this spacetime or the plane-front properties of the gravitational wave. 
It holds for any weak-field geometry of spacetime as was shown by Synge
\cite{Synge:1960}.

\subsection{Metric in optical coordinates}
\label{sect:indMetricOC}

To obtain the metric in optical coordinates, we need to invert 
the coordinate transformation rules, equations (\ref{OC:barU}),
(\ref{OC:barX}), (\ref{OC:barY}), (\ref{OC:barV}), and substitute the 
resulting formulae in the fundamental form (\ref{metricForm2}). We can 
then group together all terms containing the same binomial 
$\rmd \bar{x}^{\mu} \rmd \bar{x}^{\nu}$ and thus obtain the
components of the induced metric $\bar{g}_{\mu\nu}$. The resulting
formulae are somewhat complicated. However, one can easily recognize 
in them the components of the metric in Fermi normal coordinates. 
Therefore, the result can be presented as a sum:
\begin{equation}
   \bar{g}_{\mu\nu} = \eta_{\mu\nu} + C_{\mu\nu} + D_{\mu\nu} ,
\end{equation}
where $C_{\mu\nu}$ has the same structure as the metric in the FN
coordinates and $D_{\mu\nu}$ is the new part.

Consider the definition for functions $P_a$ and $Q_a$ given by 
(\ref{defPAlong}) and (\ref{defQAlong}). Using the fact that   
$u-s \approx \bar{z} +\bar{r}$ which is valid to first order in $h$, we 
obtain the formulae for these functions in optical coordinates:
\begin{eqnarray}
   P_a & = & h_a + \hat{h}_a - 2 f_a , \label{defPA_OC}\\
   Q_a & = & h_a - \frac{1}{2} (\bar{z} + \bar{r}) \, \hat{h}'_a  - 
             f_a . \label{defQA_OC}
\end{eqnarray}
Then the first part of the metric in optical coordinates can be written as
\begin{eqnarray}
   C_{xx} & = & P_{+} , \\
   C_{yy} & = & - P_{+} , \\
   C_{xy} & = & P_{\times} , \\
   C_{xz} & = & - \frac{1}{\bar{z} + \bar{r}} \left( 
                    \bar{x} P_{+} + \bar{y} P_{\times} \right), \\
   C_{yz} & = & - \frac{1}{\bar{z} + \bar{r}} \left( 
                    \bar{x} P_{\times} - \bar{y} P_{+} \right), \\
   C_{zz} & = & \frac{1}{(\bar{z} + \bar{r})^2} \left[ 
                    (\bar{x}^2 - \bar{y}^2) P_{+} + 
                    2 \bar{x} \bar{y} P_{\times} \right] ,\\
   C_{\tau x} & = & - \frac{1}{\bar{z} + \bar{r}} \left( 
                        \bar{x} Q_{+} + \bar{y} Q_{\times} \right), \\
   C_{\tau y} & = & - \frac{1}{\bar{z} + \bar{r}} \left( 
                        \bar{x} Q_{\times} - \bar{y} Q_{+} \right), \\
   C_{\tau z} & = & \frac{1}{(\bar{z} + \bar{r})^2} \left[  
                        (\bar{x}^2 - \bar{y}^2) Q_{+} + 
                        2 \bar{x} \bar{y} Q_{\times} \right], \\
   C_{\tau\tau} & = & \frac{1}{(\bar{z} + \bar{r})^2} \left[ 
                       (\bar{x}^2 - \bar{y}^2) (2 Q_{+} - P_{+}) + 
                       2 \bar{x} \bar{y} (2 Q_{\times} - P_{\times}) \right] .
\end{eqnarray}
If we note that $\bar{z} + \bar{r} = u - s$ in optical coordinates, and that 
$\bar{z} = u - s$ in Fermi normal coordinates, the formulae for $C_{\mu \nu}$ 
are identical in both coordinate systems.

Next, we introduce two new quantities: $R$ and $S$ according to the 
following definitions: 
\begin{eqnarray}
   R & = &  \bar{x} (Q_{+} - P_{+}) + \bar{y} (Q_{\times} - P_{\times}) ,\\
   S & = &  \bar{x} (Q_{\times} - P_{\times}) - \bar{y} (Q_{+} - P_{+}) .  
\end{eqnarray}
Then the second part of the metric can be written as
\begin{eqnarray}
   D_{xx}    & = & \frac{2}{\bar{r} (\bar{z} + \bar{r})} \bar{x} R, \\
   D_{yy}    & = & \frac{2}{\bar{r} (\bar{z} + \bar{r})} \bar{y} S, \\
   D_{xy}    & = & \frac{1}{\bar{r} (\bar{z} + \bar{r})} 
                             \left( \bar{x} S + \bar{y} R \right) , \\
   D_{xz}    & = & \frac{1}{\bar{r} (\bar{z} + \bar{r})^2} \left[ 
                           \bar{z} (\bar{z} +  \bar{r}) R - \bar{x} 
                                   (\bar{x} R + \bar{y} S) \right], \\
   D_{yz}    & = & \frac{1}{\bar{r} (\bar{z} + \bar{r})^2} \left[ 
                           \bar{z} (\bar{z} +  \bar{r}) S - \bar{y} 
                                   (\bar{x} R + \bar{y} S) \right], \\
   D_{zz}    & = & - \frac{2}{\bar{r} (\bar{z} + \bar{r})^2} 
                       \bar{z} \left( \bar{x} R + \bar{y} S \right), \\
   D_{\tau x} & = & - \frac{1}{\bar{r} (\bar{z} + \bar{r})^2} 
                       \bar{x} \left( \bar{x} R + \bar{y} S \right), \\
   D_{\tau y} & = & - \frac{1}{\bar{r} (\bar{z} + \bar{r})^2} 
                       \bar{y} \left( \bar{x} R + \bar{y} S \right), \\
   D_{\tau z} & = & - \frac{1}{\bar{r} (\bar{z} + \bar{r})^2} 
                       \bar{z} \left( \bar{x} R + \bar{y} S \right), \\
   D_{\tau\tau} & = & 0 .
\end{eqnarray}
Optical coordinates are a lesser known alternative to Fermi normal 
coordinates for analysis of a gravitational wave from the point of
view of an inertial observer.

\section{Wave synchronous coordinates}
\label{sect:waveSync}

In Sections \ref{sect:fermiNorm} and \ref{sect:optical} we solved the 
boundary-value problem using the main (non-singular) solution for the
connecting geodesic. As we know, the geodesic equation can also have
the singular solution, for which 
\begin{equation}\label{WS:pvo}
   p_{v0} = 0 .
\end{equation}
We will now consider the normal coordinate construction based on the 
singular solution for the connecting geodesic. In this case, we have to 
abandon the Fermi condition of orthogonality (\ref{FN:cutOrtho})
because it is not compatible with (\ref{WS:pvo}).

\begin{figure}[t]
   \centering\includegraphics[width=0.75\textwidth]{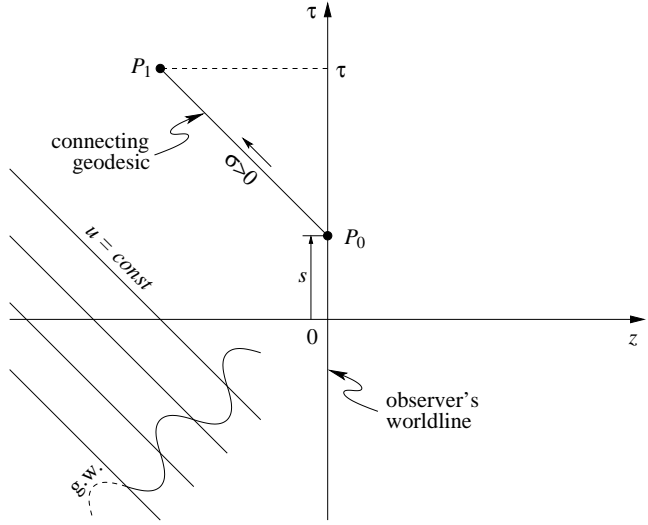}
   \caption{The observer's worldline $z=0$ and the connecting
     geodesic $P_0 P_1$ in the boundary-value problem for
     wave-synchronous coordinates. $P_0 P_1$ lies in the plane $u =
     {\mathrm{const}}$ which coincides with one of the surfaces of
     constant phase of the gravitational wave. Negative values of 
     parameter $\sigma$ correspond to the second connecting
     geodesic $P_0 P_2$, where $P_2 =\{\tau, -x, -y, z \}$. (In
     $z\tau$ plane $P_2$ appears at the same location as $P_1$.)}
   \label{fig:wavesync}
\end{figure}

We take the singular solution described in 
Section~\ref{sect:geodesic2}. Then $u$ is constant along the geodesic:
\begin{equation}\label{WS:u=u0}
   u(\sigma) = u_0 .
\end{equation}
In other words, advancing parameter $\sigma$ makes the corresponding
point $x^{\mu}(\sigma)$ move along the geodesic in such a way that 
it remains fixed to a particular phase front of the gravitational
wave, as shown in figure~\ref{fig:wavesync}. 
For this reason, we will call the normal coordinates that are based on this
solution wave-synchronous coordinates.

As before, we assume that the connecting geodesic originates from the 
worldline of the observer, at point $P_0=\{s,0,0,0\}$, and ends at an
arbitrary point in this spacetime, $P_1=\{\tau,x,y,z\}$. The affine 
parameter along the geodesic takes the value $\sigma = 0$ at
point $P_0$ and $\sigma = \sigma_1$ at point $P_1$. The solution for 
the geodesic equation (Section~\ref{sect:geodesic2}) 
connects the coordinates of points $P_0$ and $P_1$: 
\begin{eqnarray}
   x & = & p_{x0} \, \sigma_1 \, (1 - h_{+}) - 
           p_{y0} \, \sigma_1 \, h_{\times} , \label{xSigmaWS} \\
   y & = & p_{y0} \, \sigma_1 \, (1 + h_{+}) - 
           p_{x0} \, \sigma_1 \, h_{\times} , \label{ySigmaWS} \\
   z    & = &     A \sigma_1^2 + B \sigma_1 , \label{zViaAB} \\
   \tau & = & s - A \sigma_1^2 - B \sigma_1 , \label{tViaAB}
\end{eqnarray}
where $h_a$ is constant along the geodesic:
\begin{equation}
   h_a = h_a(u) = h_a(u_0) .
\end{equation}
The boundary value problem is to determine arbitrary constants 
$p_{x0}, p_{y0}, A, B, s$, $\sigma_1$ in terms of the coordinates of
point $P_1$.

First, we find parameter $s$ a function of the coordinates of $P_1$. 
By definition, $s = u_0$. Since $u$ is constant along the geodesic 
(\ref{WS:u=u0}), we find that
\begin{equation}
   s = u, \qquad {\mathrm{or}} \qquad s = \tau + z .
\end{equation}
Second, inverting (\ref{xSigmaWS}) and (\ref{ySigmaWS}) to first order
in $h$, we obtain
\begin{eqnarray}
   p_{x0} & = & (1 + h_{+}) \frac{x}{\sigma_1} + 
              h_{\times} \, \frac{y}{\sigma_1} , \label{px0WS} \\
   p_{y0} & = & (1 - h_{+}) \frac{y}{\sigma_1} + 
              h_{\times} \, \frac{x}{\sigma_1} . \label{py0WS}
\end{eqnarray}
Substituting these equations in the normalization condition 
(\ref{normPexplicit}) with $\epsilon = 1$, we obtain 
the formula for $\sigma_1$:
\begin{equation}
   \sigma_1^2 = x^2 + y^2 + (x^2 - y^2) h_{+} + 2 x y \, h_{\times} .
\end{equation}
Naturally, there are two solutions:
\begin{equation}
   \sigma_1 = \pm \left[ x^2 + y^2 + (x^2 - y^2) h_{+} + 
              2 x y \, h_{\times}  \right]^{1/2} .
\end{equation}
The solution with the ``$-$'' sign corresponds to another singular
geodesic with the same parameter $u_0$. This second geodesic connects 
$P_0$ with point $P_2 =\{\tau, -x, -y, z \}$ which is the inverse of 
point $P_1$ in the $xy$-plane. Therefore, we can safely 
discard this solution. Taking the ``$+$'' sign and keeping only terms
first order in $h$, we obtain the final solution for $\sigma_1$: 
\begin{equation}
   \sigma_1 \approx \rho + \frac{1}{2\rho} \left[ 
      (x^2 - y^2) \, h_{+} + 2 x y \, h_{\times} \right] , 
\end{equation}
where $\rho = \sqrt{x^2 + y^2}$.

Next, we can substitute $p_{x0}$ and $p_{y0}$ in (\ref{geodPuWS})
and find the constant $A$ to first order in $h$: 
\begin{equation}
   A = \frac{1}{4 \sigma_1^2} \left[ (x^2 - y^2) h'_{+} + 
      2 x y \, h'_{\times} \right] .
\end{equation}
Knowing $A$ and $\sigma_1$, we can find $B$ from (\ref{zViaAB}),
\begin{equation}
   B = \frac{1}{\sigma_1} \left( z - A \sigma_1^2 \right) .
\end{equation}
It follows from equations (\ref{zViaAB}) and (\ref{tViaAB}) that 
\begin{equation}\label{pt_pz_at_0}
   \left. p^{z}    \right|_{\sigma = 0} = - 
   \left. p^{\tau} \right|_{\sigma = 0} = B .
\end{equation}

We now turn to the derivation of the coordinate transformation rules
using the definition (\ref{defNormCoord}). Consider first the $x$
coordinate: 
\begin{eqnarray}
   \bar{x} & \equiv & \lambda^{\bar{x} x} p_{x0} \sigma_1 + 
          \lambda^{\bar{x} y} p_{y0} \sigma_1 \nonumber \\
    & = & x + \frac{1}{2} x \, h_{+} + \frac{1}{2} y \, h_{\times} .
\end{eqnarray}
Similarly, we derive the transformation rule for the $y$ coordinate:
\begin{eqnarray}
   \bar{y}  & \equiv & \lambda^{\bar{y} y} p_{y0} \sigma_1 + 
          \lambda^{\bar{y} x} p_{x0} \sigma_1 \nonumber \\
    & = & y - \frac{1}{2} y \, h_{+} + \frac{1}{2} x \, h_{\times} .
\end{eqnarray}
Using (\ref{pt_pz_at_0}), we can obtain a formula for $z$,
\begin{eqnarray}
   \bar{z} & \equiv & \lambda^{\bar{z}}{}_{z} \left. p^{z} 
       \right|_{\sigma=0} \sigma_1 \nonumber \\
    & = & B \sigma_1 \nonumber \\
    & = & z - A \sigma_1^2 ,
\end{eqnarray}
and for $\tau$,
\begin{eqnarray}
   \bar{\tau} & \equiv & s + \lambda^{\bar{\tau}}{}_{\tau} \left. p^{\tau} 
       \right|_{\sigma=0} \sigma_1 \nonumber \\
    & = & s - B \sigma_1 \nonumber \\
    & = & \tau + A \sigma_1^2 .
\end{eqnarray}
We have thus obtained the transformation rules for wave-synchronous
coordinates, which can be summarized as
\begin{eqnarray}
   \bar{x} & = & x + \frac{1}{2} x \, h_{+}(u) + 
                     \frac{1}{2} y \, h_{\times}(u) , 
                           \label{XinWS} \\
   \bar{y} & = & y - \frac{1}{2} y \, h_{+}(u) + 
                     \frac{1}{2} x \, h_{\times}(u) , 
                           \label{YinWS} \\
   \bar{z} & = & z       - \frac{1}{4} (x^2 - y^2) h'_{+}(u) -  
                           \frac{1}{2} x y h'_{\times}(u) , 
                           \label{ZinWS} \\
   \bar{\tau} & = & \tau + \frac{1}{4} (x^2 - y^2) h'_{+}(u) +
                           \frac{1}{2} x y h'_{\times}(u) .
                           \label{TinWS}
\end{eqnarray}
Equivalently, one can use $u,v$ instead of $z,\tau$: 
\begin{eqnarray}
   \bar{u} & = & u , \label{UinWS} \\
   \bar{v} & = & v + \frac{1}{2} (x^2 - y^2) h'_{+}(u) + x y h'_{\times}(u) .
                     \label{VinWS}
\end{eqnarray}

To find the metric in wave-synchronous coordinates we need to invert
the coordinate transformation rules, equations 
(\ref{XinWS}), (\ref{YinWS}), (\ref{UinWS}), (\ref{VinWS}) and
substitute the resulting formulae in the fundamental form 
(\ref{metricForm2}). The result is
\begin{equation}\label{fundFormWS}
   F = - \rmd \bar{u} \, \rmd \bar{v} + \rmd \bar{x}^2 + 
         \rmd \bar{y}^2 - 2 \Phi \rmd \bar{u}^2 ,
\end{equation}
where $\Phi$ is given by 
\begin{equation}
   \Phi = - \frac{1}{4} (\bar{x}^2 - \bar{y}^2) \, h''_{+}(\bar{u})
          - \frac{1}{2} \bar{x} \bar{y} \, h''_{\times}(\bar{u}) .
\end{equation}
Therefore, the nonzero components of the metric in wave-synchronous 
coordinates are 
\begin{eqnarray}
  \bar{g}_{xx} & = & 1 ,                \label{WSmetricGxx} \\
  \bar{g}_{yy} & = & 1 ,                \label{WSmetricGyy} \\
  \bar{g}_{zz} & = & 1 - 2 \Phi ,       \label{WSmetricGzz} \\
  \bar{g}_{\tau z} & = & - 2 \Phi ,      \label{WSmetricGtz} \\
  \bar{g}_{\tau \tau} & = & - 1 - 2 \Phi .\label{WSmetricGtt}
\end{eqnarray}
This metric represents the exact solution of the Einstein equations 
found by Peres \cite{Peres:1959} and also by Ehlers and Kundt 
\cite{Ehlers:1962}. The equivalence of the exact solution and the
metric of the wave-synchronous coordinates was previously 
discussed in \cite{Rakhmanov:2005}.

We have seen that Fermi normal coordinates are valid for arbitrarily
large distances in the longitudinal ($\bar{z}$) direction but are
limited in the transverse ($\bar{x} \bar{y}$) plane. Since 
wave-synchronous coordinates correspond to the exact solution of 
general relativity, they are not restricted in space the way Fermi 
normal coordinates are. The exact solution requires special attention 
in the case when $|\Phi| \sim 1$ for which $\bar{g}_{\tau \tau}$ may
vanish. Analysis of the metric in this regime belongs to the study of 
exact solutions of general relativity and is outside the scope of this
paper.

\section{Comparison of the coordinate systems}
\label{sect:physics}

In this Section, we briefly consider how the different coordinate
systems can be used to describe the motion of an inertial test mass
which is placed in the field of a gravitational wave. In what follows,  
the normal coordinates will appear as $x, y, z, \tau$, i.e. without
the overline, for simplicity.

\subsection{TT gauge versus normal coordinates}

As we have seen in Section~\ref{sect:observer}, an inertial test mass
that is initially at rest in the TT coordinate system will remain at 
rest even in the presence of a gravitational wave. Here the words 
``at rest'' only mean that the coordinates of the test mass are not
changing. One way to realize such a coordinate system would be to use 
inertial masses themselves to define the coordinate grid. Imagine a 
large number of inertial masses in space forming a 3-dimensional cubic 
lattice and assume that initially no mass is moving relative to the
other. For any point on the grid, its order numbers along the three 
lattice dimensions would yield the TT coordinates of that location. 
Assume that a test mass is introduced in this space, and it is at
rest with respect to this grid. Then there will be no relative motion 
between the test mass and the co-located mass on the grid even in the 
presence of a gravitational wave. Therefore, the coordinates of the
test mass will not be changing. However, the proper distance between 
any two masses on the grid will be changing due to changes in the 
metric that are caused by the gravitational wave. An effort to make 
changes in the proper distance between two masses appear as changes 
in their coordinates would lead to normal coordinate construction. 
In a normal coordinate system, an inertial test mass will appear to 
be moving with respect to the coordinate grid, i.e. its coordinates 
will be changing. The details of this motion will depend on the type 
of the normal coordinates used. We consider here the motion of an
inertial test mass under the influence of a gravitational wave for two
normal coordinate systems described above: Fermi and wave-synchronous.

\subsection{Motion of a test mass in Fermi normal coordinates}

If the test mass was at rest in TT coordinates, it will be moving in 
FN coordinates. The exact dependence of the test mass coordinates on
time is given by equations (\ref{FNequivX})--(\ref{FNequivT}) 
in which we have to assume that the TT coordinates are constant.  
In FN coordinates, the test mass is experiencing acceleration
which indicates the presence of forces produced by
the gravitational wave. To analyze these forces we need to find
explicit formulae for the acceleration of the test mass.  
Differentiating equations (\ref{FNequivX})--(\ref{FNequivT})
twice with respect to time and replacing the constant TT
coordinates with their FN counterparts, we obtain
\begin{eqnarray}
   \frac{\rmd^2 x}{\rmd \tau^2} & = & 
        \frac{1}{2} x \, h''_{+} +
        \frac{1}{2} y \, h''_{\times}  + z
        \left( x \, H''_{+} + y H''_{\times} \right) ,
        \label{physFN_ddx} \\
   \frac{\rmd^2 y}{\rmd \tau^2} & = &
        \frac{1}{2} x \, h''_{\times} - 
        \frac{1}{2} y \, h''_{+}  + z 
        \left( x \, H''_{\times} - y \, H''_{+} \right) , 
        \label{physFN_ddy} \\
   \frac{\rmd^2 z}{\rmd \tau^2} & = & -
        \frac{1}{2} (x^2 - y^2) H''_{+} - x y H''_{\times} ,
        \label{physFN_ddz}
\end{eqnarray}
where primes denote derivatives with respect to $\tau$. The function 
$H_a$ was introduced in (\ref{defHfunct}) and its derivatives 
\begin{eqnarray}
   H'_a  & = & \frac{1}{z^2} \left[  h_a(\tau + z) - 
                  h_a(\tau) - z h'_a(\tau) \right] , 
               \label{diffHa} \\
   H''_a & = & \frac{1}{z^2} \left[  h'_a(\tau + z) - 
                  h'_a(\tau) - z h''_a(\tau) \right] ,
\end{eqnarray}
are finite in the limit of $z \rightarrow 0$.

One can also find the test mass acceleration directly in FN 
coordinates, bypassing the TT gauge altogether. Indeed, taking 
the equations for a geodesic in FN coordinates,
\begin{equation}\label{accelinFN}
   \frac{\rmd^2 x^{i}}{\rmd \tau^2} = \frac{1}{2} 
      \frac{\partial C_{\tau\tau}}{\partial x^{i}} - 
      \frac{\partial C_{\tau i} }{\partial \tau} ,  
      \qquad {\mathrm{for}} \qquad i = 1, 2, 3 ,
\end{equation}
and substituting in them the formulae for the metric coefficients 
$C_{\tau i}$ and $C_{\tau \tau}$ from  
(\ref{FNmetricGtx})--(\ref{FNmetricGtt}) we can obtain equations  
(\ref{physFN_ddx})--(\ref{physFN_ddz}). These equations describe the
acceleration of the test mass in response to the propagating
gravitational wave. In this form, they are not easy to
interpret. We know for example, that the longitudinal acceleration
(along $z$) is much smaller than the acceleration in the transverse 
directions ($x, y$) but this is not obvious from these equations.

Interpretation of the acceleration will be straightforward 
if the equations are presented in the Newtonian form. This can be
done as follows. First, we replace $\tau$ with $t$ via 
$\tau = c t$. Then we introduce the scalar field $\phi$ 
and the vector field ${\mathbf{b}}$ according to the definitions: 
\begin{eqnarray}
   C_{\tau \tau} & = & - \frac{2}{c^2} \phi ,\\
   C_{\tau i}    & = & - \frac{1}{c^2} b_i .
\end{eqnarray}
With these notational changes, equations (\ref{accelinFN}) can be
written in the Newtonian form:
\begin{equation}\label{NewtonEq}
   \frac{\rmd^2 {\mathbf{r}}}{\rmd t^2} = - \nabla \phi + 
      \frac{1}{c} \frac{\partial {\mathbf{b}}}{\partial t} . 
\end{equation}
Here $\phi$ represents the dominant part of the acceleration 
produced by the gravitational wave and ${\mathbf{b}}$ generates 
relativistic corrections. (The relativistic corrections come from 
both $\phi$ and ${\mathbf{b}}$, but only $\phi$ generates
the non-relativistic part.)

Explicit formula for the scalar field can be found from 
(\ref{FNmetricGtt}), 
\begin{equation}\label{phi_FN_exactPQ}
   \phi = - \frac{c^2}{2 z^2} \left[ (x^2 - y^2) 
                \left( 2 Q_{+} - P_{+} \right) + 2 x y 
                \left( 2 Q_{\times} - P_{\times} \right) \right] . 
\end{equation}
Explicit formulae for the vector field can be found from 
(\ref{FNmetricGtx}) -- (\ref{FNmetricGtz}),
\begin{eqnarray}
   b_{x} & = & \frac{c^2}{z} \left( x Q_{+} + y Q_{\times} \right) ,\\
   b_{y} & = & \frac{c^2}{z} \left( x Q_{\times} - y Q_{+} \right) ,\\
   b_{z} & = & - \frac{c^2}{z^2} \left[  
                 (x^2 - y^2) Q_{+} + 2 x y Q_{\times} \right] .
\end{eqnarray}
Note that equation (\ref{phi_FN_exactPQ}) can also be written as
\begin{equation}\label{phi_FN_exact}
   \phi = - \frac{c^2}{2} \left[ (x^2 - y^2) H'_{+} + 
                2 x y H'_{\times} \right] ,
\end{equation}
where $H_a'$ is given by (\ref{diffHa}).

The non-relativistic approximation of (\ref{NewtonEq}) can be obtained 
by expanding scalar field $\phi$ and vector field ${\mathbf{b}}$ in 
powers of $1/c$. The leading terms in this
expansion are given by the second-order approximation for the metric, 
(\ref{2orderGtx})--(\ref{2orderGtt}). We will also need to
make the substitution:
\begin{equation}\label{defEta}
   h_a(\tau + z) = \eta_a \left( t + \frac{z}{c} \right) ,
\end{equation}
which will allow expansion of $h_a$ in powers of $1/c$. Consistency
requires that we expand $\phi$ to the first order:
\begin{equation}\label{phi0plus1}
   \phi = \phi_0 + \phi_1 ,
\end{equation}
where $\phi_0$ is purely non-relativistic and $\phi_1$ is of order
$1/c$. Consider first the non-relativistic part of the potential 
$\phi_0$. Taking the leading (zeroth-order) terms in the expansion of 
$H_a$ in (\ref{phi_FN_exact}) we find that 
\begin{equation}\label{phiFN_nonrel}
   \phi_0 =  - \frac{1}{4} (x^2 - y^2) \ddot{\eta}_{+}(t)
             - \frac{1}{2} x y \, \ddot{\eta}_{\times}(t) .
\end{equation}
This part defines the non-relativistic approximation for the 
test mass acceleration:
\begin{eqnarray}
  - \frac{\partial \phi_0}{\partial x} & = & 
                   \frac{1}{2} x \, \ddot{\eta}_{+}(t)
                 + \frac{1}{2} y \, \ddot{\eta}_{\times}(t) ,
                   \label{accelXphi} \\
  - \frac{\partial \phi_0}{\partial y} & = & 
                   \frac{1}{2} x \, \ddot{\eta}_{\times}(t)
                 - \frac{1}{2} y \, \ddot{\eta}_{+}(t) ,
                   \label{accelYphi} \\
  - \frac{\partial \phi_0}{\partial z} & = & 0 .
                   \label{accelZphi}
\end{eqnarray}
In this picture, the gravitational wave manifests itself through the
time-dependent potential which generates the forces acting on the mass 
(one for each polarization $+$ and $\times$). These forces are
orthogonal to the direction of the gravitational-wave propagation. 
Therefore, the motion of the test mass caused by these forces is 
confined to the transverse plane.

In the non-relativistic approximation, the scalar field satisfies 
the Laplace equation 
\begin{equation}\label{Laplace}
   \nabla^2 \phi_0 = 0 .
\end{equation}
The same equation is satisfied by gravitational potentials away from
the sources in Newtonian physics. In this regard, the scalar field 
can be viewed as a special case of a gravitational potential.

The relativistic corrections come from $\phi_1$ and
${\mathbf{b}}$. Taking the first-order terms in the expansion of $H_a$ in 
(\ref{phi_FN_exact}) we find that 
\begin{equation}\label{phi_FNorder1}
   \phi_1 = - \frac{1}{12 c} z \left[ (x^2 - y^2) {\eta}^{(3)}_{+}(t)
            + 2 x y \, {\eta}^{(3)}_{\times}(t) \right] . 
\end{equation}
Expanding (\ref{2orderGtx})--(\ref{2orderGtz}) in powers of $1/c$ and
keeping the leading order terms, we obtain 
\begin{eqnarray}
   b_x & = & \frac{1}{3} z \left[ x \, \ddot{\eta}_{+}(t)  
                                + y \, \ddot{\eta}_{\times}(t) \right] , 
             \label{approxBx} \\
   b_y & = & \frac{1}{3} z \left[ x \, \ddot{\eta}_{\times}(t)
                                - y \, \ddot{\eta}_{+} (t) \right] ,
             \label{approxBy} \\
   b_z & = & - \frac{1}{3} \left[ (x^2 - y^2) \ddot{\eta}_{+}(t) 
                   + 2 x y \, \ddot{\eta}_{\times}(t) \right] .
             \label{approxBz} 
\end{eqnarray}
Thus, the first-order relativistic correction to the test mass
acceleration is given by
\begin{eqnarray}
  - \frac{\partial \phi_1}{\partial x} 
  + \frac{1}{c} \frac{\partial b_x}{\partial t} & = & 
    \frac{1}{2c} z \left[ x \, \eta^{(3)}_{+}(t)
                        + y \, \eta^{(3)}_{\times}(t) \right] ,
                    \label{accelXRel} \\
  - \frac{\partial \phi_1}{\partial y}
  + \frac{1}{c} \frac{\partial b_y}{\partial t} & = & 
    \frac{1}{2c} z \left[ x \, \eta^{(3)}_{\times}(t)
                        - y \, \eta^{(3)}_{+}(t) \right] ,
                    \label{accelYRel} \\
  - \frac{\partial \phi_1}{\partial z}
  + \frac{1}{c} \frac{\partial b_z}{\partial t} & = & 
  - \frac{1}{4 c} \left[ (x^2 - y^2) \, \eta^{(3)}_{+}(t)
                    + 2 x y \, \eta^{(3)}_{\times}(t) \right] .
                        \label{accelZRel}
\end{eqnarray}

We conclude by noting that vector field ${\mathbf{b}}$ in the
non-relativistic approximation, (\ref{approxBx})--(\ref{approxBz}),
is divergence free,
\begin{equation}\label{divB}
   \nabla \cdot {\mathbf{b}} = 0 .
\end{equation}
This means that there is a vector field ${\mathbf{a}}$ such that 
\begin{equation}
   {\mathbf{b}} = \nabla \times {\mathbf{a}} .
\end{equation}
The definition of ${\mathbf{a}}$ is not unique. 
Here we give one possible realization for this vector field: 
\begin{eqnarray}
   a_x & = & \frac{1}{3} x^2 y \, \ddot{\eta}_{+}(t)  
           + \frac{1}{6} x y^2 \, \ddot{\eta}_{\times}(t) , \\
   a_y & = & \frac{1}{3} x y^2 \, \ddot{\eta}_{+}(t) 
           - \frac{1}{6} x^2 y \, \ddot{\eta}_{\times} (t) , \\
   a_z & = & \frac{1}{3} x y z \, \ddot{\eta}_{+}(t) 
           - \frac{1}{6} z (x^2 - y^2) \ddot{\eta}_{\times}(t) ,
\end{eqnarray}
which can be useful for calculations of the motion of continuous media
(fluids or elastic bodies) in the presence of a gravitational wave.

Note that the dependence of the potential $\phi$ on $z$ and $t$ is not 
consistent with the relativistic form of the gravitational-wave 
propagation, e.g. (\ref{defEta}). The same is true for the vector 
field ${\mathbf{b}}$ in which $z$ and $t$ are not related in any
way. This problem does not occur in wave-synchronous coordinate
system, as we will see next.

\subsection{Motion of a test mass in wave-synchronous coordinates}

The description of test mass motion in wave-synchronous coordinates 
is similar to the description in Fermi normal coordinates. In 
wave-synchronous coordinates, the test mass will be 
moving under the influence of the gravitational wave. The exact 
dependence of the test mass coordinates on time is given by equations 
(\ref{XinWS})--(\ref{ZinWS}) in which we have assume that the 
TT coordinates are constant. To analyze the forces acting on the test
mass we need to find the explicit formulae for its acceleration.  
Differentiating equations (\ref{XinWS})--(\ref{ZinWS}) twice 
with respect to time and replacing the constant TT
coordinates with their wave-synchronous counterparts, we obtain
\begin{eqnarray}
   \frac{\rmd^2 x}{\rmd \tau^2} & = & 
      \frac{1}{2} x \, h''_{+}(u) +
      \frac{1}{2} y \, h''_{\times}(u) ,\label{physWS_x} \\
   \frac{\rmd^2 y}{\rmd \tau^2} & = & 
      \frac{1}{2} x \, h''_{\times}(u) -
      \frac{1}{2} y \, h''_{+}(u) , \label{physWS_y} \\
   \frac{\rmd^2 z}{\rmd \tau^2} & = & - 
      \frac{1}{4} (x^2 - y^2) h'''_{+}(u) -
      \frac{1}{2} x y \, h'''_{\times}(u) ,
      \label{physWS_z}
\end{eqnarray}
where primes denote derivatives with respect to $u$.

One can find the test mass acceleration directly in 
wave-synchronous coordinates, bypassing the TT gauge altogether. 
Indeed, taking the equations for a geodesic in wave-synchronous
coordinates \cite{Rakhmanov:2005},
\begin{equation}\label{accelinWS}
   \frac{\rmd^2 x^{i}}{\rmd \tau^2} = \frac{1}{2} 
      \frac{\partial g_{\tau\tau}}{\partial x^{i}} - 
      \frac{\partial g_{\tau i} }{\partial \tau} ,
      \qquad {\mathrm{for}} \qquad i = 1, 2, 3,
\end{equation}
and substituting in them the formulae for the metric coefficients 
$g_{\tau i}$ and $g_{\tau \tau}$ from 
(\ref{WSmetricGtz}) and (\ref{WSmetricGtt}), we can obtain equations  
(\ref{physWS_x})--(\ref{physWS_z}).

By replacing $\tau$ with $t$ and introducing the scalar and vector
fields, 
\begin{eqnarray}
   1 + g_{\tau \tau} & = & - \frac{2}{c^2} \phi ,\\
   g_{\tau i} & = & - \frac{1}{c^2} b_i ,
\end{eqnarray}
we can present equations (\ref{accelinWS}) in the Newtonian form:  
\begin{equation}
   \frac{\rmd^2 {\mathbf{r}}}{\rmd t^2} = - \nabla \phi + 
      \frac{1}{c} \frac{\partial {\mathbf{b}}}{\partial t} .
\end{equation}
Here $\phi$ represents the dominant part of the acceleration 
produced by the gravitational wave and ${\mathbf{b}}$ generates 
relativistic corrections. 
In the explicit form, the scalar field is given by 
\begin{equation}\label{phi_WS}
   \phi = - \frac{1}{4} (x^2 - y^2) \,
            \ddot{\eta}_{+}\left( t + \frac{z}{c} \right) 
          - \frac{1}{2} x y \, 
            \ddot{\eta}_{\times} \left( t + \frac{z}{c} \right) .
\end{equation}
The vector field has the following components:
\begin{eqnarray}
   b_x & = & 0, \\
   b_y & = & 0, \\
   b_z & = & 2 \phi .
\end{eqnarray}
The scalar field $\phi$ plays the role of the potential which
generates the forces acting on the test mass. Note that in 
wave-synchronous coordinates the potential acquires the full 
$z$-dependence consistent with the relativistic nature of the 
gravitational wave. This can also be seen from the fact that 
the potential satisfies the wave equation: 
\begin{equation}
   \nabla^2 \phi = \frac{1}{c^2} \frac{\partial^2 \phi}{\partial t^2}.
\end{equation}
One can think of this potential as the fully relativistic version of
the potential in Fermi normal coordinates.

Consider now the vector field ${\mathbf{b}}$. In wave-synchronous
coordinates ${\mathbf{b}}$ produces a purely longitudinal acceleration
of the test mass. In fact, it is equal to twice the acceleration
produced by the potential (in the opposite direction): 
\begin{equation}
   \frac{1}{c} \frac{\partial b_z}{\partial t} = 
     2 \frac{\partial \phi}{\partial z} .
\end{equation}
Also note that vector field ${\mathbf{b}}$ is not divergence free. 
Instead, it satisfies
\begin{equation}
   \nabla \cdot {\mathbf{b}} = \frac{2}{c} 
      \frac{\partial \phi}{\partial t}.
\end{equation}
This equation can be written in the covariant 4-dimensional form:
\begin{equation}
   \partial_{\mu} b^{\mu} \equiv - \partial_{\tau} b_{\tau} + 
      \nabla \cdot {\mathbf{b}} = 0 ,
\end{equation}
where 4-vector $b_{\mu}$ is formed from ${\mathbf{b}}$ by adding 
the time component, $b_{\tau} = 2 \phi$. Therefore, 
\begin{equation}
   b_{\mu} = \{ 2 \phi, 0, 0, 2 \phi \} .
\end{equation}
Another way to introduce the 4-vector $b_{\mu}$ is
through the metric tensor:
\begin{equation}
   g_{\tau \mu} = \eta_{\tau \mu} - \frac{1}{c^2} b_{\mu} ,
\end{equation}
where $g_{\tau \mu}$ are the components of the metric in
wave-synchronous coordinates.

The non-relativistic approximation for the test mass acceleration can 
be found by expanding (\ref{phi_WS}) in powers of $1/c$. It is
interesting to note that the zeroth and the first order approximations
for the acceleration in wave-synchronous coordinates are identical to
those in Fermi normal coordinates. The advantage of wave-synchronous
coordinate system is that it allows us to have the acceleration to all
orders in $1/c$ and provides a mathematical formalism which is fully 
consistent with the relativistic nature of the gravitational wave.

\section{Conclusion}
\label{sect:conclu}

Normal coordinates are a convenient tool for analysis of the effects 
of gravitational waves from the point of view of an inertial
observer. We have revisited the normal-coordinate construction 
for a plane gravitational wave and showed that it depends on the 
boundary-value problem for the connecting geodesic. Three different 
types of the boundary-value problem have been considered in this
paper. The first is based on a non-singular spacelike connecting 
geodesic that is orthogonal to the observer's worldline. This 
construction leads to Fermi normal coordinates. The second 
boundary-value problem is based on a null connecting geodesic and
leads to optical coordinates. The third is based on a singular
spacelike connecting geodesic and leads to wave-synchronous
coordinates. For each type of the boundary-value problem we obtained 
explicit formulae for the coordinate transformation rules and the 
induced metric. These formulae are exact as long as the calculations 
stay within the linearized theory of gravitation. In particular, 
they are valid for arbitrarily large distances in the longitudinal
direction. Also, we showed that the exact formulae yield the
infinite-series representation for Fermi normal coordinates and the
induced metric. We have thus found that the infinite series for FN 
coordinates and the induced metric can actually be summed and the
result of this summation can be given in a closed analytical form.

Historically, Fermi-normal coordinates have always been limited to 
the long-wavelength regime. However, as we have shown, Fermi normal 
coordinates can actually be defined far beyond the long-wavelength
regime. No change in point of view and no transition of any kind 
occurs at distances equal to or comparable with the wavelength of 
the gravitational wave. We can now conclude that the range limitation 
that has always been associated with Fermi normal coordinates was 
rather unnecessary. Without this limitation, Fermi normal coordinates 
become a viable alternative to the TT coordinates for theoretical 
studies of gravitational waves. We also found that wave-synchronous 
coordinates yield the exact solution of Peres and Ehlers-Kundt. Since 
this solution is globally defined, the system of wave-synchronous 
coordinates is valid for arbitrarily large distances. This was
possible due to the special geometry of spacetime which represents 
a plane gravitational wave propagating in a flat background.

\ack

I would like to thank many people with whom I discussed this work,
particularly Kip Thorne who stressed the importance of these
calculations at the beginning of this research, and also Joe Romano
and the late Leonid Grishchuk with whom I had many interesting
conversations about the role of the coordinate system in the detector
response near the end of this work. In addition, I would like to thank 
Rick Savage and Joe Romano for comments on the final version of the
paper, and Anton Gribovskiy for verifying the calculations of the 
induced metric. This paper took a long time to mature. The work
started at the Center for Gravitational Wave Physics at Penn State, 
continued at the LIGO Hanford Observatory, and was finished at the 
Center for Gravitational Wave Astronomy at the University of Texas at 
Brownsville. At Penn State, this research was supported by the US 
National Science Foundation under grants PHY 00-99559, 02-44902,
03-26281, and 06-00953. At LIGO, it was supported by a Visiting 
Research Fellowship from Caltech. At Brownsville, it was supported by 
the US National Science Foundation under grants HRD 0734800 and
1242090, and by the US NASA University Research Centers under grant 
NNX 09AV06A. The first version of this paper was circulated within 
the LIGO Scientific Collaboration in 2006. This paper has been
assigned LIGO Document Number P060066.

\section*{References}

\providecommand{\newblock}{}

\appendix

\section{Christoffel coefficients and Riemann tensor}
\label{app:Christoff-Riemann}

In the linearized theory of gravitation Christoffel coefficients are
given by
\begin{equation}\label{deffChystoff}
   \Gamma_{\alpha\beta\gamma} = \frac{1}{2} \left( 
        h_{\alpha\beta,\gamma} + h_{\alpha\gamma,\beta} - 
        h_{\beta\gamma,\alpha} \right) .
\end{equation}
For the metric defined in (\ref{metricForm2}),  
$h_{\alpha\beta}$ depends essentially on one coordinate:
\begin{equation}
   h_{\alpha\beta} = h_{\alpha\beta}(u) , 
      \qquad {\mathrm{where}} \qquad 
   u = \tau + z .
\end{equation}
Then the non-zero independent Christoffel coefficients are  
\begin{eqnarray}
   & & \Gamma_{xxu} = -\Gamma_{uxx} = - \Gamma_{yyu} = \Gamma_{uyy} = 
       \frac{1}{2} \, h'_{+}(u) ,\\
   & & \Gamma_{xyu} = - \Gamma_{uxy} = \Gamma_{yxu} = 
      \frac{1}{2} \, h'_{\times}(u) ,
\end{eqnarray}
where primes stand for differentiation with respect to $u$. 
Switching from $u, v$ to $\tau, z$, we find  
\begin{eqnarray}
   & & \Gamma_{xx\tau} = -\Gamma_{\tau xx} = \Gamma_{xxz} = -\Gamma_{zxx} = 
           \frac{1}{2} \, h'_{+}(\tau + z) , \\
   & & \Gamma_{yy\tau} = -\Gamma_{\tau yy} =  \Gamma_{yyz} = -\Gamma_{zyy} = 
          -\frac{1}{2} \, h'_{+}(\tau + z) ,\\
   & & \Gamma_{xy\tau} = -\Gamma_{\tau xy} = \Gamma_{yx\tau} = 
           \frac{1}{2} \, h'_{\times}(\tau + z) ,\\
   & & \Gamma_{xyz} = -\Gamma_{zxy} = \Gamma_{yxz} = 
           \frac{1}{2} \, h'_{\times}(\tau + z) ,
\end{eqnarray}
where primes stand for differentiation with respect to $z$ or $\tau$.

The components of the Riemann tensor \cite{Misner:1973} are defined
according to 
\begin{equation}
   R^{\mu}{}_{\nu\alpha\beta} = \Gamma^{\mu}{}_{\nu\beta,\alpha} - 
      \Gamma^{\mu}{}_{\nu\alpha,\beta} + 
      \Gamma^{\mu}{}_{\rho\alpha} \Gamma^{\rho}{}_{\nu\beta} -
      \Gamma^{\mu}{}_{\rho\beta}  \Gamma^{\rho}{}_{\nu\alpha} .
\end{equation}
In the linearized theory, this definition reduces to
\begin{equation}
   R_{\mu\nu\alpha\beta} =
      \Gamma_{\mu\nu\beta,\alpha} - \Gamma_{\mu\nu\alpha,\beta} .
\end{equation}
Substituting for Christoffel coefficients from (\ref{deffChystoff}),
we obtain 
\begin{equation}
   R_{\mu\nu\alpha\beta} = \frac{1}{2} 
      ( h_{\mu\beta,\nu\alpha} - h_{\mu\alpha,\nu\beta} +  
        h_{\nu\alpha,\mu\beta} - h_{\nu\beta,\mu\alpha} ).
\end{equation}
There are only three non-zero independent components of the Riemann
tensor corresponding to metric defined in (\ref{metricForm2}):
\begin{eqnarray}
   & & R_{xuxu} = R_{yuyu} = - \frac{1}{2} \, h''_{+}(u) ,\\
   & & R_{xuyu}= - \frac{1}{2} \, h''_{\times}(u) .
\end{eqnarray}
Switching from $u, v$ to $\tau, z$, we find 
\begin{eqnarray}
   & & R_{x \tau x \tau} = R_{x \tau x z} = R_{x z x z} = 
             - \frac{1}{2} \, h''_{+}(\tau + z) ,\\
   & & R_{y \tau y \tau} = R_{y \tau y z} = R_{y z y z} = 
               \frac{1}{2} \, h''_{+}(\tau + z) ,\\
   & & R_{x \tau y \tau} = R_{x \tau y z} = R_{x z y \tau} = R_{x z y z} = 
             - \frac{1}{2} \, h''_{\times}(\tau + z) .
\end{eqnarray}

\end{document}